# Inconsistency Robustness for Logic Programs

**Carl Hewitt**

***This article is dedicated to Alonzo Church and Stanisław Jaśkowski***

## Abstract

This article explores the role of Inconsistency Robustness in the history and theory of Logic Programs. Inconsistency Robustness has been a continually recurring issue in Logic Programs from the beginning including Church's system developed in the early 1930s based on *partial* functions (defined in the lambda calculus) that he thought would allow development of a general logic without the kind of paradoxes that had plagued earlier efforts by Frege, *etc.*[1]

Planner [Hewitt 1969, 1971] was a kind of hybrid between the procedural and logical paradigms in that it featured a procedural embedding of logical sentences in that an implication of the form (*p* ***implies*** *q*) can be procedurally embedded in the following ways:

Forward chaining

- **When asserted** *p*, **Assert** *q*
- **When asserted** ¬*q*, **Assert** ¬*p*

Backward chaining

- **When goal** *q*, **SetGoal** *p*
- **When goal** ¬*p*, **SetGoal** ¬*q*

Developments by different research groups in the fall of 1972 gave rise to a controversy over Logic Programs that persists to this day in the form of following alternatives:

1. **Procedural Interpretation:** Logic Programs using *procedural interpretation of logic-clause syntax for a program*[i] [Kowalski 2014]
2. **Procedural Embedding:** Logic Program in which *each computational step*[ii] *is logically inferred*[iii]

[i] a Logic Program is written as $\Psi \Leftarrow (\Phi_1 \wedge \ldots \wedge \Phi_n)$, which is logically equivalent to the disjunctive clause $\Psi \vee \neg\Phi_1 \vee \ldots \vee \neg\Phi_n$ where $\Psi$ and each of the $\Phi_i$ is either $P[t_1, \ldots, t_m]$ or $\neg P[t_1, \ldots, t_m]$ for some atomic predicate P and terms $t_j$).

[ii] *e.g.,* using the Actor Model of computation

[iii] *e.g.,* using Direct Logic

This article argues for the second alternative based on the following considerations:

- Programs in logic-clause syntax are a special case of the second alternative because each computational step of a program in logic-clause syntax is logically inferred as backward chaining or forward chaining.
- Reducing propositions to logic-clause syntax can obscure their natural structure.
- Procedural interpretation of logic-clause syntax can obscure the natural structure of proofs.[i]

**Logic-clause syntax is far too limited to be of use for general Logic Programs.**

Kowalski advocates a bold thesis: "*Looking back on our early discoveries, I value most the discovery that computation could be subsumed by deduction.*"[2] (Roman numeral superscripts in text are endnotes at the end of this article.)

However, mathematical logic cannot always infer computational steps because computational systems make use of arbitration for determining which message is processed next. Since reception orders are in general indeterminate, they cannot be inferred from prior information by mathematical logic alone. Therefore mathematical logic alone cannot in general implement computation. Logic Programs (like Functional Programs) are useful idioms even though they are not universal. For example Logic Programs can provide useful principles and methods for systems which are quasi-commutative and quasi-monotonic even though the systems themselves cannot be implemented using Logic Programs.

A fundamental principle of Inconsistency Robustness is to make contradictions explicit so that arguments for and against propositions can be formalized. In the Inconsistency Robustness paradigm, deriving contradictions is a progressive development and contradictions are not "game stoppers" that they would be using classical logic (in which reasoning about inconsistent information can make erroneous inferences). Contradictions can be helpful instead of being something to be "swept under the rug" by denying their existence or fruitlessly attempting complete elimination in systems of practice that are pervasively inconsistent.[3]

A contradiction is manifest when both a proposition and its negation are asserted even if by different parties, *e.g.*, New York Times said "*Snowden is a whistleblower.*", but NSA said "*Snowden is not a whistleblower.*"[4]

---

[i] *e.g.,* Natural Deduction using Direct Logic

This paper explores the role of Inconsistency Robustness in the development and theory of Logic Programs, which is an interesting test case involving pervasive inconsistency in an area in which, traditionally, inconsistency was not supposed to occur.

| **Inconsistency Robustness** |
| --- |
| In this article, boxes like this one are used below to call out instances of inconsistency robustness. |

## Uniform Proof Procedures based on Resolution

> *By this it appears how necessary it is for nay man that aspires to true knowledge to examine the definitions of former authors; and either to correct them, where they are negligently set down, or to make them himself. For the errors of definitions multiply themselves, according as the reckoning proceeds, and lead men into absurdities, which at last they see, but cannot avoid, without reckoning anew from the beginning; in which lies the foundation of their errors...*
> [Hobbes *Leviathan*, Chapter 4][5]

An important limitation of classical logic for inconsistent theories is that it supports the principle that from an inconsistency anything and everything can be inferred, *e.g. "The moon is made of green cheese."*

For convenience, I have given the above principle the name IGOR[6] for **I**nconsistency in **G**arbage **O**ut **R**edux. IGOR can be formalized as follows in which a contradiction about a proposition **Φ** infers any proposition **Ψ**,[i] *i.e.*, $\mathbf{\Phi, \neg\Phi \vdash \Psi}$.

**Of course, IGOR *cannot* be part of inconsistency-robust logic because it allows *every* proposition to be inferred from a contradiction.**

The IGOR principle of classical logic may not seem very intuitive! So why is it included in classical logic?

- **Classical Proof by Contradiction:** $(\mathbf{\Psi} \vdash \mathbf{\Phi, \neg\Phi}) \Rightarrow (\vdash \neg\mathbf{\Psi})$, which can be justified in classical logic on the grounds that if **Ψ** infers a contradiction in a consistent theory then **Ψ** must be false. In an inconsistent theory. Classical Proof by Contradiction leads to explosion by the following derivation in classical logic by a which a contradiction about P infers any proposition **Θ**:

$$\underline{P, \neg P} \vdash \neg\underline{\Theta \vdash P, \neg P} \vdash (\neg\neg\Theta) \vdash \Theta$$

[i] Using the symbol ⊢ to mean "infers in classical mathematical logic." The symbol was first published in [Frege 1879].

- **Classical Contrapositive for Inference:** $(\Psi \vdash \Phi) \Rightarrow (\neg\Phi \vdash \neg\Psi)$, which can be justified in classical logic on the grounds that if $\Psi \vdash \Phi$, then if **Φ** is false then **Ψ** must be false. In an inconsistent theory. Classical Contrapositive for Inference leads to explosion by the following derivation in classical logic by a which a contradiction about P (*i.e.,* $\vdash P, \neg P$ **)** infers any proposition **Θ** by the following proof:

  Since $\vdash P, \neg\Theta \vdash P$ by monotonicity. Therefore $\neg P \vdash \Theta$ by Classical Contrapositive for Inference. Consequently $P, \neg P \vdash \Theta$**.**
- **Classical Extraneous ∨ Introduction**:[7] $\Psi \vdash (\Psi\vee\Phi)$, which in classical logic says that if Ψ is true then $\Psi\vee\Phi$ is true regardless of whether Φ is true.[8] In an inconsistent theory, Extraneous ∨ introduction leads to explosion via the following derivation in classical logic in which a contraction about P infers any proposition **Θ**:

  $$\underline{P, \neg P} \vdash \underline{(P\vee\Theta), \neg P} \vdash \Theta$$
- **Classical Excluded Middle:** $\vdash (\Psi\vee\neg\Psi)$, which in classical logic says that $\Psi\vee\neg\Psi$ is true regardless of whether Ψ is true. *Excluded Middle* is the principle of Classical Logic that for every proposition X the following holds: ExcludedMiddle[X] ≡ $X\vee\neg X$

  However, Excluded Middle is not suitable for inconsistency-robust logic because it is equivalent[i] to saying that there are no inconsistencies, *i.e.,* for every proposition X,

  Noncontradiction[X] ≡ $\neg(X\wedge\neg X)$

  Using propositional equivalences, note that

  ExcludedMiddle[$\Phi\vee\Psi$] ⇔ $(\Psi\vee\neg\Psi\vee\Phi)\wedge(\Phi\vee\neg\Phi\vee\Psi)$

  Consequently, ExcludedMiddle[$\Phi\vee\Psi$]⇒$(\Psi\vee\neg\Psi\vee\Phi)$, which means that the principle of Excluded Middle implies $\Psi\vee\neg\Psi\vee\Phi$ for all propositions Ψ and Φ. Thus the principle of Excluded Middle is not inconsistency robust because it implies every proposition Φ can be proved[ii] given any contradiction Ψ. [Kao 2011]

---

[i] using propositional equivalences

[ii] using ∨-*Elimination*, *i.e.*, $\neg\Phi, (\Phi\vee\Psi) \vdash_T \Psi$

**Classical Logic is unsafe for use with potentially inconsistent information.**[i]

[Robinson 1965] developed a deduction method called resolution which was proposed as a uniform proof procedure using first-order logic for proving theorems which

> *Converted everything to clausal form and then used a method analogous to modus ponens to attempt to obtain a contradiction by adding the negation of the proposition to be proved.*

Resolution uniform proof procedures were used to generate some simple proofs [Wos, et. al. 1965; Green 1969; Waldinger and Lee 1969; Anderson and Bledsoe 1970; *etc.*]. In the resolution uniform proof procedure theorem proving paradigm, the use of procedural knowledge was considered to be "*cheating.*"[9]

Kowalski (also see [Decker 2003, 2005, 2010]) advocates that inference for inconsistent information systems[10] be performed using resolution theorem proving.[11] Unfortunately, in the presence of inconsistent information, resolution theorem provers can prove propositions using invalid arguments.

---

[i] Turing noted that classical logic can be used to make invalid inferences using inconsistent information "*without actually going through* [an explicit] *contradiction.*" [Diamond 1976] Furthermore, [Church 1935, Turing 1936] proved that it is computationally undecidable whether a mathematical theory of practice is inconsistent.

For example, consider the following illustration:[i]

> A resolution theorem prover proved $\Psi\vee\Phi$ and as a consequence it was added to the information system. Sometime afterward, another resolution theorem prover proved $\Phi$ making use of the previously proved $\Psi\vee\Phi$**.** Unfortunately, $\Phi$ was proved by resolution theorem proving using an invalid argument because[12]
>
> - The first resolution theorem prover proved $\Psi\vee\Phi$ because[ii] $\vdash\Psi$[iii] where $\Psi$ holding has no bearing[iv] on whether $\Phi$ holds.[v]
> - Using an invalid argument,[vi] the second resolution theorem proved $\Phi$ using the previously proved $\Psi\vee\Phi$ because[vii] $\vdash \neg\Psi$[viii].

The above illustration illustrates how resolution theorem proving can prove propositions[ix] using invalid arguments.

For a large information base $\Omega$ of theories of practice (*e.g.,* a climate models, theories of the human brain, etc.), Kowalski advocates use of the resolution set of support strategy[13] as follows: In proving a proposition $\Phi$, the set of support strategy chooses resolutions involving resolving $\neg\Phi$ with $\Omega$ and resolving the resulting clauses with each other and with $\neg\Phi$.[14]

---

[i] The illustration below is presented in general terms for simplicity of exposition because presenting a detailed example involves significant complexity with lots of irrelevant details. An important point is that resolution theorem provers rely on overlap of vocabulary in deciding on which propositions to resolve. Consequently, they can become entangled in irrelevancy when deriving a contradiction when two propositions (having no bearing on whether the other holds) share vocabulary.

[ii] unknown to the resolution theorem prover

[iii] even though the resolution theorem prover could *not* prove $\Psi$ in the time that it took to prove $\Psi\vee\Phi$

[iv] unknown to the resolution theorem prover although $\Psi$ and $\Phi$ have some common vocabulary, which can lead a resolution theorem prover astray

[v] Unfortunately, the following rule of classical logic is not inconsistency robust: $\Psi\vdash\Psi\vee\Phi$. However, a resolution theorem prove must honor *all* of the rules of classical logic.

[vi] because the resolution theorem provers unknowingly used $\vdash\Psi,\neg\Psi$ as the basis for the proof of $\Phi$ although $\Psi$ holding has no bearing on $\Phi$ holding even though they have some common vocabulary

[vii] unknown to the resolution theorem prover

[viii] even though the resolution theorem prover could *not* prove $\neg\Psi$ in the time that it took to prove $\Phi$

[ix] for pervasively inconsistent information systems of practice, *e.g.*, theories for climate change and the human brain

Any proposition X can be proved[i] by interacting with a resolution theorem prover[ii] from an information base containing propositions $\vartheta$ and $\neg\vartheta$[iii] as follows:

1) $\vartheta \vdash_{\mathit{SetOfSupport}} (\vartheta \vee X)$ because $\vartheta, \underline{\neg(\vartheta \vee X)} \vdash_{\mathit{ResolutionTheoremProver}}$ False[iv]
2) $\vartheta \vee X, \neg\vartheta \vdash_{\mathit{SetOfSupport}} X$ because
   $\vartheta \vee X, \neg\vartheta, \underline{\neg X} \vdash_{\mathit{ResolutionTheoremProver}}$ False[v]
3) A resolution theorem prover *using set of support* working on an information base containing both $\vartheta$ and $\neg\vartheta$ can be used to first prove $\vartheta \vee X$ using **1)** and then can be used to prove X using **2)**.

Direct Logic was developed to address limitations[15] of classical logic as a foundation for Computer Science:

- Classical Direct Logic is intended to be thought to be consistent by an overwhelming consensus of working professional mathematicians. See [Hewitt 2013] for discussion of Classical Direct Logic.
- Inconsistency Robust Direct Logic is for pervasively inconsistent theories of practice, e.g., theories for climate modeling and for modeling the human brain. Classical Direct Logic can be freely used in theories of Inconsistency Robust Direct Logic. See [Hewitt 2010] for discussion of Inconsistency Robust Direct Logic.

Inconsistency Robust Direct Logic is the state of the art for possibly inconsistent information.[vi] However, enforcing the constraint that a resolution theorem prover use only the rules of Inconsistency Robust Direct Logic is inefficient as well as extraordinarily complicated and awkward. Consequently, resolution theorem provers are unsuitable for inconsistency robust reasoning.

---

[i] using an invalid argument because it used inconsistent assumptions

[ii] using the set of support strategy

[iii] a resolution theorem prover using set of support might not realize that $\vartheta$ and $\neg\vartheta$ are inconsistent because if they are not in the set of support, they will not be resolved and it might not be immediately apparent from their clausal forms that they are inconsistent.

[iv] set of support is underlined

[v] set of support is underlined

[vi] Many other paraconsistent logics have been proposed that are partially inconsistency robust.

**Inconsistency Robustness: Resolution**

Using resolution as the only rule of inference is problematical because it can obscure the natural structure of propositions and the natural structure of proofs using Natural Deduction. Also, Resolution Theorem Proving can prove propositions using invalid arguments for systems of practice because of pervasive inconsistencies.

### Procedural Embedding redux

In the late 1960's, the two major paradigms for constructing semantics software systems were procedural and logical. The procedural paradigm was epitomized by Lisp [McCarthy *et. al*. 1962] which featured recursive procedures that operated on list structures including property lists that were updated imperatively. The logical paradigm was epitomized by uniform Resolution Theorem Provers [Robinson 1965].

**Inconsistency Robustness: Uniform Proof Procedures**

Uniform proof procedures using resolution was intended be a general theorem proving paradigm. But it suffered immense inefficiency in practice. Changing an axiomatization to improve performance was considered to be "cheating."

## Planner

Planner [Hewitt 1969, 1971] was a kind of hybrid between the procedural and logical paradigms in that it featured a procedural embedding of logical sentences in that an implication of the form ($p$ ***implies*** $q$) can be procedurally embedding in the following ways:[i]

Forward chaining

- **When asserted** $p$, **Assert** $q$
- **When asserted** $\neg q$, **Assert** $\neg p$

Backward chaining

- **When goal** $q$, **SetGoal** $p$
- **When goal** $\neg p$, **SetGoal** $\neg q$

Planner was the first programming language based on using explicit assertions and goals processed using pattern-directed invocation. The development of

[i] In modern notation [see appendix of this paper]:

Forward chaining

**When** $\vdash \Psi \rightarrow \vdash \Phi$

**When** $\vdash \neg \Phi \rightarrow \vdash \neg \Psi$

Backward chaining

**When** $\Vdash \Phi \rightarrow \Vdash \Psi$

**When** $\Vdash \neg \Psi \rightarrow \Vdash \neg \Phi$

Planner was inspired by the work of Jaśkowski [1934], Newell and Simon [1956], McCarthy [1958], McCarthy *et. al.* [1962], Minsky [1958], Polya [1954], and Popper [1935, 1963].

> **Planner represented a rejection of the resolution uniform proof procedure paradigm in favor of Procedural Embedding making use of new program language constructs[16] for computation including using explicit assertions and goals that invoked programs.**

A subset called Micro-Planner was implemented by Gerry Sussman, Eugene Charniak and Terry Winograd as an extension to Lisp primarily for pragmatic reasons since it saved memory space and processing time (both of which were scarce) by comparison with more general problem solving techniques, *e.g.,* [Polya 1957]:

- Lisp was well suited to the implementation of a Micro-Planner interpreter.
- The full functionality of Lisp libraries were immediately available for use by Micro-Planner programs.
- The Lisp compiler could be used to compile Lisp programs used by Micro-Planner applications to make them smaller and run faster. (It was unnecessary to first implement a Micro-Planner compiler.)

Computers were expensive. They had only a single slow processor and their memories were very small by comparison with today. So Planner adopted some efficiency expedients including the following:[i]

- Backtracking [Golomb and Baumert 1965] was adopted to economize on the use of time and storage by working on and storing only one possibility at a time in exploring alternatives. In several ways, backtracking proved unwieldy helping to fuel the great control structure debate. Hewitt investigated some preliminary alternatives in his thesis.
- A unique name assumption was adopted by assuming that different names referred to different objects, which saved space and time. For example names like Peking and Beijing were assumed to refer to different objects.
- A closed world assumption could be implemented by conditionally testing whether an attempt to prove a goal exhaustively failed. Later this capability was given the misleading name "negation as failure" because for a goal G it was possible to say: "if attempting to achieve G exhaustively fails then assert (*Not* G)."[ii]
- Being a hybrid language, Micro Planner had two different syntaxes, variable binding mechanisms, etc. So it lacked a certain degree of elegance. In fact, after Hewitt's lecture at IJCAI'71, Allen Newell rose

---

[i] Prolog later also adopted these same efficiency expedients.

[ii] satirized as "*the less that can proved, the more that can be assumed!*"

from the audience to remark on the lack of elegance in the language! However, variants of this syntax have persisted to the present day.

Micro-Planner was used in Winograd's natural-language understanding program SHRDLU [Winograd 1971], Eugene Charniak's story understanding work, work on legal reasoning [McCarty 1977], and some other projects. This generated a great deal of excitement in the field of Artificial Intelligence.

**Inconsistency Robustness: Procedural Embedding**

Planner was designed as a program language for Procedural Embedding. However, efficiency expedients were made in its implementation that unfortunately resulted in inflexible problem solving strategies as well as awkward, limited reasoning capabilities.

Although Winograd made an impressive demo, the successors of Planner and SHRDLU were incapable of practically realizing Procedural Embedding because of limited hardware performance and lack of effective software frameworks and tooling. Because of decades of subsequent progress, it has become feasible to developed practical, principled systems for Procedural Embedding.

Logic Programs in ActorScript are a further development of Planner. For example, suppose there is a grounded-complete predicate[17] Street[*aName, anAddress, anotherAddress, aDistance*] that is true exactly when the street with *aName*, *anAddress* and *anotherAddress* has *aDistance*.

**When** ⊩ Street[*aName*, *anAddress*, *anAddress*, *aDistance*]→

// when a goal is set for a distance between *anAddress* and itself

⊢ *aDistance*=0▮ // assert that the distance from an address to itself is 0

The following goal-driven Logic Program works forward from *start* to find the distance to *finish*:

**When** ⊩ Distance[*start*, *finish*, *aDistance*]→

⊢ *aDistance*=Minimum.[{*nextDistance* + *remainingDistance*

| ⊩ Street[*aName*, *start*, *next*, *nextDistance*],

Distance[*next*, *finish*, *remainingDistance*]}[18]]▮

// the distance from *start* to *finish* is the minimum of the set of the sums of the

// distance for the next address after *start* and

// the distance from that address to *finish*

The following goal-driven Logic Program works backward from *finish* to find the distance from *start*:

**When** ⊩ Distance[*start*, *finish*, *aDistance*]→

⊢ *aDistance*=

Minimum.[{*remainingDistance* + *previousDistance*

| ⊩ Street[*aName*, *previous*, *finish*, *previousDistance*],

Distance[*start*, *previous*, *remainingDistance*]}[19]]▮

// the distance from *start* to *finish* is the minimum of the set of the sums of the

// distance for the previous address before *finish* and

// the distance from *start* to that address

Note that all of the above Logic Programs work together concurrently providing information to each other.

The following procedure computes the shortest path from one location to another:

ShortestPath.[start, finish] ≡

start �

finish ⸮ [start], // the shortest path from start to itself is [start]

**else** ⸮

[start, // the shortest path begins with start

∨ ⊩ Distance[start,

*next*, // the next location on the shortest path has

Minimum.[{*aDistance* // the minimum distance to finish

| ⊩ Street[*anAddress*, start, *aLocation*≠start, _]

Distance[*aLocation*, finish, *aDistance*]}]]

**once**→ // only need one location for the next in shortest path

ShortestPath.[*next*, finish]]▮ // the shortest path continues with *next*

## Control Structure Controversies

Peter Landin introduced a powerful co-routine control structure using his **J** (for Jump) operator that could perform a nonlocal goto into the middle of a procedure invocation [Landin 1965]. In fact the **J** operator enabled a program to jump back into the middle of a procedure invocation even after it had already returned!

Drew McDermott and Gerry Sussman called Landin's concept "*Hairy Control Structure*" and used it in the form of a nonlocal goto for the Conniver program language [McDermott and Sussman 1972]. Hewitt and others were skeptical about hairy control structure. Pat Hayes [1974] remarked: *Their* [Sussman and McDermott] *solution, to give the user access to the implementation primitives of Planner, is however, something of a retrograde step (what are Conniver's semantics?)*

The difficulties using backtracking in Planner and Conniver were useful in that they provoked further research into control structures for procedural embedding.

**Inconsistency Robustness: Control Structures**

There was there germ of a good idea (previously emphasized in Polya [1957] and "progressive deepening" [de Groot 1965]) in Conniver; namely, using co-routines to computationally shift focus to another branch of investigation while keeping alive the one that has been left Scott Fahlman used this capability of Conniver to good effect in his planning system for robot construction tasks [Fahlman 1973] to introduce a set of higher-level control and communications operations for its domain. However, the ability to jump back into the middle of procedure invocations seemed awkward and confusing. Hairy Control Structure didn't seem to be what was needed as the foundation to solve the difficulties in communication that were a root cause of the control structure difficulties.

## Control structures are patterns of passing messages

In November 1972. Alan Kay visited MIT and gave an inspiring lecture that explained some of his ideas for Smalltalk-72 building on the message-passing of Planner, Simula [Dahl and Nygaard 1967] as well as the Logo work of Seymour Papert with the "little person" model of computation used for teaching children to program (*cf.* [Whalley 2006]).

The Actor model [Hewitt, Bishop, and Steiger 1973] was a new model of computation that differed from previous models of computation in that it was grounded by the laws of physics so that it could be completely general in terms of control structure.[20] It took some time to develop practical program languages for the Actor model.

Work on Planner was temporarily suspended in favor of intensive investigation of the Actor model.[i]

**Inconsistency Robustness: Message Passing**

Planner aimed to extend what could be programmed using logical methods but did not take a stand about the theoretical limits of these methods. However, once the Actor model was invented in late 1972, it became clear that logical inference alone would not suffice for computation because the order of Actor message reception could not always be logically inferred.

> [Hewitt 1976] reported
> *... we have found that we can do without the paraphernalia of "hairy control structure" (such as possibility lists, non-local gotos, and assignments of values to the internal variables of other procedures in CONNIVER.)...* ***The conventions of ordinary message-passing seem to provide a better structured, more intuitive foundation for constructing the communication systems needed for expert problem-solving modules to cooperate effectively.*** (emphasis in original)

### *Edinburgh Logic for Computable Functions*

Like Planner,[21] Edinburgh Logic for Computable Functions [Milner 1972; Gordon, Milner, and Wadsworth 1979] was capable of both forward chaining as well as backward chaining. This was accomplished by a purely functional program operating on a special data type called "Theorem" to produce new theorems by forward and backward chaining. Sub-goaling strategies (called tactics) were represented as higher-order functions taking strategies as arguments and returning them as results with goal failure implemented using exceptions.

**Inconsistency Robustness: Logic for Computable Functions**

Edinburgh Logic for Computable Functions was a notable advance in that its logical soundness was guaranteed by the type system. However, its problem solving generality is limited since it was not concurrent because it was purely functional.

[i] Work on Logic Programs later resumed in the Scientific Community Model [see section below].

## Procedural Embedding versus Procedural Interpretation of Logic-clause Syntax

At Edinburgh, Pat Hayes and Bob Kowalski collaborated on resolution theorem proving. Then Hayes visited Stanford where Bruce Baumgart published his *Micro-Planner Alternate Reference Manual* in April 1972. Hayes says that from the time that he learned about Micro-Planner it seemed obvious to him that it was based on controlled deduction.[22]

When he returned to Edinburgh, he talked about his insight with anyone who would listen and gave internal seminars at two of the major departments at Edinburgh concerned with logic. In the third department, Hayes point seemed irrelevant because they were busy getting their hands on the latest "magic machinery" for controlling reasoning using Popler [Davies 1973], a derivative of Planner. Hayes wrote a joint paper with Bruce Anderson on "*The Logicians Folly*" against the resolution uniform proof procedure paradigm [Anderson and Hayes 1972].

Gerry Sussman and Seymour Papert visited Edinburgh spreading the news about Micro-Planner and SHRDLU casting doubt on the resolution uniform proof procedure approach that had been the mainstay of the Edinburgh Logicists. According to Maarten van Emden [2006]

> *The run-up to the workshop* [Machine Intelligence 6 organized by Donald Michie in 1970] *was enlivened by telegrams from Seymour Papert at MIT announcing on alternating days that he was (was not) coming to deliver his paper entitled "The Irrelevance of Resolution", a situation that caused Michie to mutter something about the relevance of irresolution. The upshot was that a student named Gerry Sussman appeared at the appointed time. It looked as if this was going to be his first talk outside MIT. His nervousness was compounded by the fact that he had been instructed to go into the very bastion of resolution theorem proving and tell the assembled experts how totally misguided they were in trying to get anything relevant to AI with their chosen approach.*
>
> *I had only the vaguest idea what all this was about. For me theorem proving was one of the things that some people (including Kowalski) did, and I was there for the programming. If Bob and I had anything in common, it was search. Accordingly I skipped the historic Sussman lecture and arrived late for the talk scheduled to come after Sussman's. Instead, I found an unknown gentleman lecturing from a seat in the audience in, what I thought a very English voice. It turned out that a taxi from the airport had delivered Seymour Papert after all, just in time for the end of Sussman's lecture, which was now being re-done properly by the man himself.*
>
> *The effect on the resolution people in Edinburgh of this frontal assault was traumatic. For nobody more so than for Bob Kowalski. Of course there was no shortage of counter objections, and the ad hoc creations of*

*MIT were not a pretty sight. But the occasion hit hard because there was a sense that these barbarians had a point.*

The above developments generated tension among the Logicists at Edinburgh. These tensions were exacerbated when the UK Science Research Council commissioned Sir James Lighthill to write a report on the AI research situation. The resulting report [Lighthill 1973; McCarthy 1973] was highly critical although SHRDLU [Winograd 1971] was favorably mentioned. "*Resolution theorem-proving was demoted from a hot topic to a relic of the misguided past. Bob* [Kowalski] *doggedly stuck to his faith in the potential of resolution theorem proving. He carefully studied Planner*." [Bruynooghe, Pereira, Siekmann, and van Emden 2004]

van Emden [2006] recalled:

> *Kowalski's apparent research program narrowed to showing that the failings so far of resolution inference were not inherent in the basic mechanism. He took great pains to carefully study PLANNER and CONNIVER. And painful it was. One of the features of the MIT work was that it assumed the audience consisted of LISP programmers. For anybody outside this circle (Kowalski most definitely was not a LISP programmer), the flavour is repellent.*

According to [Kowalski 2014]

> *Pat Hayes and I had been working in Edinburgh on a book [Hayes and Kowalski,1971] about resolution theorem-proving, when he returned from a second visit to Stanford (after the first visit, during which he and John McCarthy wrote the famous situation calculus paper [McCarthy and Hayes, 1968]). He was greatly impressed by Planner, and wanted to rewrite the book to take Planner into account. I was not enthusiastic, and we spent many hours discussing and arguing about the relationship between Planner and resolution theorem proving. Eventually, we abandoned the book, unable to agree.*

In the fall of 1972 at MIT, there was universal dissatisfaction with the adequacy of micro-Planner. Fundamental work[i] proceeded long before the name "Functional Program" was introduced. Likewise, as recounted in this article, fundamental work proceeded for decades before the name "Logic Program" was introduced.

---

[i] *e.g.* [Church 1932, McCarthy *et. al*. 1962]

To further the development of Procedural Embedding, the Actor Model was invented, which provided a rigorous basis for defining both:

- *Functional Program*: Actors do not change
- *Logic Program*: Each computational step is logically inferred.

Hayes reported that he was astonished when Kowalski wrote back from Marseilles saying that he and Colmerauer had a revolutionary idea that Horn clauses could be interpreted as backward-chaining programs. Feeling that his ideas were being unfairly appropriated by Kowalski, Hayes complained to the head of their unit Bernard Meltzer and still feeling unsatisfied wrote a summary and exegesis of his ideas in a paper for the proceedings of a summer school in Czechoslovakia with the idea of recording the priority of his ideas [Hayes 1973].

However, Kowalski felt that his work with Colmerauer bore little resemblance to anything that had been discussed previously in Edinburgh by Hayes claiming that Hayes' ideas (and the paper that he published) were based on using equations for computation (in the spirit of the work in Aberdeen [Foster and Elcock 1969].

Kowalski [2008] recalled:

> *In the meanwhile, critics of the formal approach, based mainly at MIT, began to advocate procedural representations of knowledge, as superior to declarative, logic-based representations. This led to the development of the knowledge representation and problem-solving languages Planner and micro-Planner. Winograd's PhD thesis (1971), using micro-Planner to implement a natural language dialogue for a simple blocks world, was a major milestone of this approach. Research in automated theorem-proving, mainly based on resolution, went into sharp decline.*
>
> *The battlefield between the logic-based and procedural approaches moved briefly to Edinburgh during the summer of 1970 at one of the Machine Intelligence Workshops organized by Donald Michie (van Emden, 2006). At the workshop, Papert and Sussman from MIT gave talks vigorously attacking the use logic in AI, but did not present a paper for the proceedings. This created turmoil among researchers in Edinburgh working in resolution theorem-proving. However, I was not convinced that the procedural approach was so different from the SL resolution system I had been developing with Donald Kuehner (1971).*
>
> *During the next couple of years, I tried to reimplement Winograd's system in resolution logic and collaborated on this with Alain Colmerauer in Marseille. This led to the procedural interpretation of Horn clauses (Kowalski 1973/1974) and to Colmerauer's development of the programming language Prolog.*

In the fall of 1972, *Prolog* (an abbreviation for "**PRO**grammation en **LOG**ique" (French for *programming in logic*)), was developed as a subset of micro-Planner that restricted programs to Horn-clause syntax[i] using backward chaining and consequently had a simpler more uniform syntax than Planner.[23] However, the restriction to Horn-clause syntax tremendously restricted the expressive power of Prolog.

Like Planner, Prolog provided the following:

- An indexed data base of propositions (limited by Prolog to positive predicates with ground arguments) and pattern-direct backward-chaining procedures (limited by Prolog to Horn-clause syntax).
- The Unique Name Assumption, by means of which different names are assumed to refer to distinct entities, *e.g.,* Peking and Beijing are assumed to be different in order to save space and time.
- The Closed World Assumption (available and used in practice in micro-Planner to save space and time although it was not strictly required by micro-Planner).

Prolog was fundamentally different in intent from Planner as follows:

- Planner was designed for **Procedural Embedding using explicit goals and assertions processed by pattern-directed procedures**. *Correctness can be checked when an assertion is made that a proposition holds in a theory using a rule of inference.* Theories do not limit the problem solving methods that can be used.[ii]
- Prolog was designed a **backward-chaining interpretation of Horn-clause syntax**. Consequently, a goal established by a Prolog backward-chaining program was intended to be a logical consequence of the *propositional interpretation* of the Prolog program syntax *without any additional checking*. A theory expressed in Prolog limits the problem solving method that can be used to backward chaining of Horn-clause syntax programs whose propositional content is the Horn clauses[24] of the theory.

A number of Logic Program features of Micro-Planner were omitted from Prolog including the following:[25]

- Using explicit assertions processed by pattern-directed procedures (*i.e.*, “forward chaining”)
- Logical negation, *e.g.*, (not (human Socrates))
- Explicit goals and subgoals that are distinct from assertions[26]

---

[i] Horn-clause syntax for a logic program is of the form $\Psi \Leftarrow (\Phi_1 \wedge \ldots \wedge \Phi_n)$ which is logically equivalent to the disjunctive clause $\Psi \vee \neg\Phi_1 \vee \ldots \vee \neg\Phi_n$ where $\Psi$ and each of the $\Phi_i$ is $P[t_1, \ldots, t_n]$ for some atomic predicate $P$ and terms $t_i$.

[ii] *e.g.,* the full range of techniques in [Polya 1957] should be available.

**In summary, Prolog was basically a subset[i] of Planner that restricted programs to a backward chaining interpretation of Horn clauses.**

According to [Kowalski 2014]:

> "... it was widely believed that logic alone is inadequate for problem solving, and that some way of controlling the theorem-prover is needed for efficiency. Planner combined logic and control in a procedural representation that made it difficult to identify the logical component. Logic programs with SLD resolution also combine logic and control, but make it possible to read the same program both logically and procedurally. I later expressed this as Algorithm=Logic+Control (A=L+C) [Kowalski, 1979a], influenced by Pat Hayes′ [1973] Computation=Controlled Deduction.
>
> The most direct implication of the equation is that, given a fixed logical representation L, different algorithms can be obtained by applying different control strategies, i.e. $A_1$=L+$C_1$ and $A_2$=L+$C_2$. Pat Hayes [1973], in particular, argued that logic and control should be expressed in separate languages, with the logic component L providing a pure, declarative specification of the problem, and the control component C supplying the problem solving strategies needed for an efficient algorithm A. Moreover, he argued against the idea, expressed by $A_1$=$L_1$+C and $A_2$=$L_2$+C, of using a fixed control strategy C, as in Prolog, and formulating the logic $L_i$ the problem to obtain a desired algorithm $A_i$."

[Hayes 1974] complained that procedural interpretation was "*alien semantics*" for logical propositions that was analogous to "*... throwing out the baby and keeping the bathwater.*" In other words, Hayes was arguing for Procedural Embedding as in Planner, whereas Kowalski was arguing for fixed control structure as in resolution uniform proof procedures.

---

[i] excepting that Prolog used unification instead of the pattern matching used in Planner. However, in practice unification is often partially turned off in Prolog programs (at the potential cost of incorrect results) because unification can be expensive.

Furthermore, there are concerns about the adequacy of logic-clause[i] syntax to express Logic Programs.

| **Inconsistency Robustness: Expressiveness of Logic Programs** |
| --- |
| Procedural Embedding (starting with Planner) allows the distinction between propositions and programs:<br>• Propositional information does not have to be reformulated using logic-clause syntax (which can obscure the natural structure of propositions).<br>• Logic Programs can be more expressive and powerful when not restricted to logic-clause syntax. For example, Logic Programs in ActorScript use Natural Deduction.<br>• Procedural Embedding can implement problem solving methods that cannot be implemented using Logic Programs (see this article on "Is Computation Subsumed by Deduction?")<br>• Assertions have provenance that justify their inference in a theory. |

## Scientific Community Model[27]

Research on the Scientific Community Model [Kornfeld and Hewitt 1981, Kornfeld 1981]. involved the development of a program language named Ether that invoked procedural plans to process goals and assertions concurrently by dynamically creating new rules during program execution. Ether also addressed issues of conflict and contradiction with multiple sources of knowledge and multiple viewpoints.

The Scientific Community Model builds on the philosophy, history and sociology of science. It was originally developed building on work in the philosophy of science by Karl Popper and Imre Lakatos. In particular, it initially made use of Lakatos' work on proofs and refutations. Subsequently development has been influenced by the work of Geof Bowker, Michel Callon, Paul Feyerabend, Elihu M. Gerson, Bruno Latour, John Law, Karl Popper, Susan Leigh Star, Anselm Strauss, and Lucy Suchman.

---

[i] Logic-clause syntax for a logic program is of the form $\Psi \Leftarrow (\Phi_1 \wedge \ldots \wedge \Phi_n)$ [logically equivalent to the disjunctive clause $\Psi \vee \neg\Phi_1 \vee \ldots \vee \neg\Phi_n$] where $\Psi$ and each of the $\Phi_i$ is of the form $P[t_1, \ldots, t_n]$ or the form $\neg P[t_1, \ldots, t_n]$ for some atomic predicate $P$ and terms $t_i$. (Logic-clause syntax is a slight generalization of Horn-clause syntax in that it allows predicates to be negated.)

In particular Latour's Science in Action had great influence. In the book, Janus figures make paradoxical statements about scientific development. An important challenge for the Scientific Community Model is to reconcile these paradoxical statements.

Scientific research depends critically on monotonicity, concurrency, commutativity, and pluralism to propose, modify, support, and oppose scientific methods, practices, and theories. Scientific Communities systems have characteristics of monotonicity, concurrency, commutativity, pluralism, skepticism and provenance:

- **monotonicity**: Once something is published it cannot be undone. Scientists publish their results so they are available to all. Published work is collected and indexed in libraries. Scientists who change their mind can publish later articles contradicting earlier ones.
- **concurrency**: Scientists can work concurrently, overlapping in time and interacting with each other.
- **quasi-commutativity**: Publications can be read regardless of whether they initiate new research or become relevant to ongoing research. Scientists who become interested in a scientific question typically make an effort to find out if the answer has already been published. In addition they attempt to keep abreast of further developments as they continue their work.
- **pluralism**: Publications include heterogeneous, overlapping and possibly conflicting information. There is no central arbiter of truth in scientific communities.
- **skepticism**: Great effort is expended to test and validate current information and replace it with better information.
- **provenance**: The provenance of information is carefully tracked and recorded.

**Inconsistency Robustness: Scientific Community Model**

The above characteristics are limited in real scientific communities. Publications are sometimes lost or difficult to retrieve. Concurrency is limited by resources including personnel and funding. Sometimes it is easier to re-derive a result than to look it up. Scientists only have so much time and energy to read and try to understand the literature. Scientific fads sometimes sweep up almost everyone in a field. The order in which information is received can influence how it is processed. Sponsors can try to control scientific activities. In Ether the semantics of the kinds of activity described in this paragraph are governed by the Actor model.

Scientific research includes generating theories and processes for modifying, supporting, and opposing these theories. Karl Popper called the process "conjectures and refutations", which although expressing a core insight, has been shown to be too restrictive a characterization by the work of Michel Callon, Paul Feyerabend, Elihu M. Gerson, Mark Johnson, Thomas Kuhn, George Lakoff, Imre Lakatos, Bruno Latour, John Law, Susan Leigh Star, Anselm Strauss, Lucy Suchman, Ludwig Wittgenstein, etc.. Three basic kinds of participation in Ether are proposing, supporting, and opposing. Scientific communities are structured to support competition as well as cooperation.

These activities affect the adherence to approaches, theories, methods, etc. in scientific communities. Current adherence does not imply adherence for all future time. Later developments will modify and extend current understandings. Adherence is a local rather than a global phenomenon. No one speaks for the scientific community as a whole.

There are a number of controversies involved in the history of Logic Programs in which different researchers took contradictory positions that are addressed in following sections of this article including, “Is computation subsumed by deduction?” and “Did Prolog-style clause programs contribute to the failure of the Japanese Fifth Generation Project (ICOT)?” and “What is a Logic Program?”

## Is Computation Subsumed by Deduction?

> “*…a single formalism suffices for both logic and computation, and logic subsumes computation.*”
>
> [Kowalski 2014]

The challenge to the generality of Logic Programs as a foundation for computation was officially thrown in *The Challenge of Open Systems* [Hewitt 1985] to which [Kowalski 1988b] replied in *Logic-Based Open Systems.* This was followed up with [Hewitt and Agha 1988] in the context of the Japanese Fifth Generation Project (see section below). All of this was in opposition to Kowalski's thesis: "*Looking back on our early discoveries, I value most the discovery that computation could be subsumed by deduction.*"[28]

In concrete terms, we cannot observe the internals of the mechanism by which the reception order of messages is determined. Attempting to do so affects the results and can even push the indeterminacy elsewhere. Instead of observing the internals of arbitration processes, we await outcomes. The reason that we await outcomes is that we have no alternative because of indeterminacy. Because of indeterminacy in the physical basis of computation, no kind of deductive mathematical logic can always infer which message will be received next and the resulting computational steps. Consequently, Logic Programs can make inferences about computation but not in general implement computation. Nevertheless, Logic Programs (like Functional Programs) can be a useful idiom.

**Inconsistency Robustness: Universality of Deduction**

**Contrary to Kowalski, computation in general cannot be subsumed by deduction.** Mathematical models of computation do not determine particular computations as follows: Arbiters can be used in the implementation of the reception order of messages, which are subject to indeterminacy. Since reception orders are in general indeterminate, they cannot be inferred from prior information by mathematical logic alone. Therefore mathematical logic alone cannot implement computation in open systems.

Actor systems can perform computations that are impossible by mathematical deduction as illustrated by the following nondeterministic Actor system[i] that can compute an integer of unbounded size:

```
Unbounded ≡
 start[ ]→                              // a start message is implemented by
   Let aCounter ← Counter[ ]       // let aCounter be a new Counter
     Do ⓛaCounter▪go[ ],
                                   // send aCounter a go  message and concurrently
       ⓛaCounter▪stop[ ]
                              //return the value of sending aCounter a stop  message

Actor theCounter Counter[ ]           // theCounter is the name of this Actor
  count ≔ 0,                             // the variable count is initially 0
  continue ≔ true
  stop[ ]→ count       //  return count
             afterward continue ≔false;
                                  //  continue is false for the next message received
  go[ ]→ continue �
            True ⦂           //  if continue is True,
              Hole theCounter▪go[ ]  //  send go[ ] to theCounter after
                after count ≔ count+1     //   incrementing count
             False ⦂ Void      // if continue is False,  return Void
```

By the semantics of the Actor model of computation [Clinger 1981; Hewitt 2006], sending Unbounded a **start** message will result in computing an integer of unbounded size.

The procedure Unbounded above can be axiomatized as follows:

$$\forall[n{:}\mathrm{Integer}]\rightarrow$$
$$\exists[aRequest{:}\mathrm{Request},\ anInteger{:}\mathrm{Integer}]\rightarrow$$
$$\mathrm{Unbounded}\ \mathbf{sent}_{aRequest}\ \mathbf{start}[\ ] \Rightarrow$$
$$\mathbf{Sent}_{\mathbf{Response}_{aRequest}}\mathrm{Returned}[anInteger] \land anInteger > n$$

However, the above axiom does *not* compute any actual output! Instead the above axiom simply asserts the *existence* of unbounded outputs for **start** messages.

---

[i] using the ActorScript programming language [Hewitt 2010a]

***Theorem.*** The nondeterministic function defined by Unbounded (above) cannot be implemented by a nondeterministic Logic Program[i].

*Proof.*[29]

> *The task of a nondeterministic Logic Program* P *is to start with an initial set of axioms and prove* Output=n *for some numeral* n*. Now the set of proofs of* P *starting from initial axioms will form a tree. The branching points will correspond to the nondeterministic choice points in the program and the choices as to which rules of inference to apply. Since there are always only finitely many alternatives at each choice point, the branching factor of the tree is always finite. Now König's lemma says that if every branch of a finitary tree is finite, then so is the tree itself. In the present case this means that if every proof of* P *proves* Output=n *for some numeral* n*, then there are only finitely many proofs. So if* P *nondeterministically proves* Output=n *for every numeral* n*, it must contain a nonterminating computation in which it does not prove* Output=n *for some numeral* n.

The following arguments support unbounded nondeterminism in the Actor model [Hewitt 1985, 2006]:

- There is no bound that can be placed on how long it takes a computational circuit called an *arbiter* to settle. Arbiters are used in computers to deal with the circumstance that computer clocks operate asynchronously with input from outside, *e.g.*, keyboard input, disk access, network input, *etc.* So it could take an unbounded time for a message sent to a computer to be received and in the meantime the computer could traverse an unbounded number of states.
- Electronic mail enables unbounded nondeterminism since mail can be stored on servers indefinitely before being delivered.
- Communication links to servers on the Internet can be out of service indefinitely.

[i] the lambda calculus is a special case of Logic Programs

The following Logic Programs procedurally embed information about Unbounded:

When ⊢*aRequest* *anActor* sent *aMessage* →
// When asserted that *anActor* is sent *aRequest* with *aMessage*
⊢*aRequest* *anActor* received *aMessage* ▮ // assert that *anActor* received *aRequest*

When ⊢*aResponse* Sent *aResult* →[30]
⊢*aResponse* Received *aResult* ▮ // assert that *aResult* is received for *aResponse*

When ⊢*aRequest* Unbounded received start[ ]→
// When asserted that Unbounded is sent Start[ ]
⊢Unbounded1[*aRequest*] Counter sent [ ],[31]
⊩$\text{Response}_{\text{Unbounded1}[request]}$ Received Returned[*aCounter*:Counter]→
// Set a goal that *aCounter* is returned for the request Unbounded1[*aRequest*].
(⊢Unbounded2[*aRequest*] *aCounter* sent go[ ],
⊢Unbounded3[*aRequest*] *aCounter* sent stop[ ],
⊩$\text{Response}_{Unbounded2[request]}$ Received Returned[Void ]→
⊩$\text{Response}_{Unbounded3[request]}$ Received Returned[*anInteger*:Integer]→
⊢$\text{Response}_{aRequest}$ Returned[*anInteger*])▮

When ⊢*aRequest* Counter received [ ]→
(⊢$\text{Response}_{aRequest}$ Returned[*aCounter*],
⊢$\text{Response}_{aRequest}$ currentCount*anAccount* = 0,
⊢$\text{Response}_{aRequest}$ continue*aCounter* = True)▮

When ⊢*aRequest* *aCounter*:Counter received Stop[ ]→
(⊢$\text{Response}_{aRequest}$ Sent Returned[currentCount*aRequest*],
When ⊢*aRequest* currentCount*aCounter* = *n* →
⊢$\text{Response}_{aRequest}$ currentCount*aCounter* = *n*,
⊢$\text{Response}_{aRequest}$ continue*aCounter*= False)▮

When ⊢*aRequest* *aCounter*:Counter received go[ ]→
(⊩*aRequest* ¬continue*aCounter*→ // Set a goal that continue*aCountet* = False
(⊢$\text{Response}_{aRequest}$ Sent Returned[Void],
⊢$\text{Response}_{aRequest}$ Unchanged*aRequest* currentCount*aCounter*,
⊢$\text{Response}_{aRequest}$ Unchanged*aRequest* continue*aCountert*),
⊩*aRequest* continue*aCounter*→ // Set a goal that continue*aCounter*= True
(⊢Counter1[*aRequest*] *aCounter* sent go[ ],
⊢Counter1[*aRequest*] Left,
// Assert that the cheese has been left at Counter1[*aRequest*]
⊩*aRequest* currentCount*aCounter* = *anInteger*
⊢Counter1[*aRequest*] currentCount*aCounter* = *anInteger* + 1,
⊩$\text{Response}_{\text{Counter1}[aRequest]}$ Received Returned[*anInteger*:Integer]→
⊢$\text{Response}_{aRequest}$ Sent Returned[*anInteger*])▮

Note that the above logic programs do *not* make use of global[32] time.[i]

**Inconsistency Robustness: Modeling Change**

Direct Logic can be used to model change in a way that is physically realizable as opposed to systems that make use of physically unrealizable global time:

1. *Global States*: a computation can be represented as a global state that determines all information about the computation. It can be nondeterministic as to which will be the next global state, *e.g.*, in simulations where the global state can transition nondeterministically to the next state as a global clock advances in time, *e.g.,* Simula [Dahl and Nygaard 1967].[i]
2. *Actors*: a computation can be represented as a configuration. Information about a configuration can be indeterminate. For example, there can be arbiters that are meta-stable and messages in transit that will be delivered at some indefinite (unbounded) time.

## The Japanese 5th Generation Project (ICOT)

Beginning in the 1970's, Japan took the DRAM market (and consequently most of the integrated circuit industry) away from the previous US dominance. This was accomplished with the help of the Japanese VLSI project that was funded and coordinated in good part by the Japanese government Ministry of International Trade and Industry (MITI) [Sigurdson 1986].

***Project Inception***

MITI hoped to repeat this victory by taking over the computer industry. However, Japan had come under criticism for "copying" the US. One of the MITI goals for ICOT was to show that Japan could innovate new computer technology and not just copy the Americans.

***Trying to go all the way with Prolog-style clause programs***

ICOT tried to go all the way with Prolog-style clause programs. Kowalski later recalled "*Having advocated LP* [Logic Programs] *as a unifying foundation for computing, I was delighted with the LP* [Logic Program] *focus of the FGCS* [Fifth Generation Computer Systems] *project.*" [Fuchi,

[i] which is physically unrealizable

Kowalski, Ueda, Kahn, Chikayama, and Tick 1993] By making Prolog-style clause programs (mainly being developed outside the US) the foundation, MITI hoped that the Japanese computer industry could leapfrog the US. "*The* [ICOT] *project aimed to leapfrog over IBM, and to a new era of advanced knowledge processing applications."* [Sergot 2004]

Unfortunately, ICOT misjudged the importance of direct message passing, *e.g*., in the Actor Model [Hewitt, Bishop, and Steiger 1973] which had been developed in reaction to the limitations of Planner.

ICOT had to deal with the concurrency and consequently developed concurrent program languages based on Prolog-style clauses [Shapiro 1989] similar[33] to the above Logic Program. However, it proved difficult to implement clause-procedure invocation in these languages as efficiently as procedures in object-oriented program languages. Simula-67 originated a hierarchical class structure for objects so that message handling procedures (methods) and object instance variables could be inherited by subclasses. Ole-Johan Dahl [1967] invented a powerful compiler technology using dispatch tables that enabled message handling procedures in subclasses of objects to be efficiently invoked. The compiler technology originally developed for Simula[34] far out-performed the ICOT compliers for Prolog-style clause languages developed by ICOT.

The clausal syntax used by ICOT was awkward because it only allowed relational syntax for procedure calls and consequently was not compositional requiring the use of multiple ports for communication.[35]

For example, below is the definition of a procedure that computes a future of a list that is the "fringe" of the leaves of tree.

```
Fringe▪[aTree]  ≡
     aTree � Leaf[x] ⦂
                    [x],   //  return a list with just the leaf terminal
                 Fork[tree1, tree2] ⦂
                    [∨Fringe▪[tree1], ∨Postpone[36] Fringe▪[tree2]]
                     //  return a list elements of the first of the left
                         //  branch followed by elements of the right branch
```

The above procedure can be used to define SameFringe that determines if two lists have the same fringe [Hewitt 1972]:

```
SameFringe▪[aTree, anotherTree] ≡
                               //  test if two trees have the same fringe
        Fringe▪[aTree] = Fringe▪[anotherTree]
```

Using clausal syntax, ICOT encountered difficulties dealing with concurrency, *e.g.*, readers-writers concurrency. Concurrency control for readers and writers in a shared resource is a classic problem. The fundamental constraint is that readers are allowed to operate concurrently but a writer is not allowed to operate concurrently with other writers and readers.

The interface for the readers/writer guardian is the same as the interface for the shared resource:

```
Interface ReadersWriter having read[Query]↦ QueryResult,
                                write[Update]↦ Void
```

State diagram of **ReadersWriter** implementations:

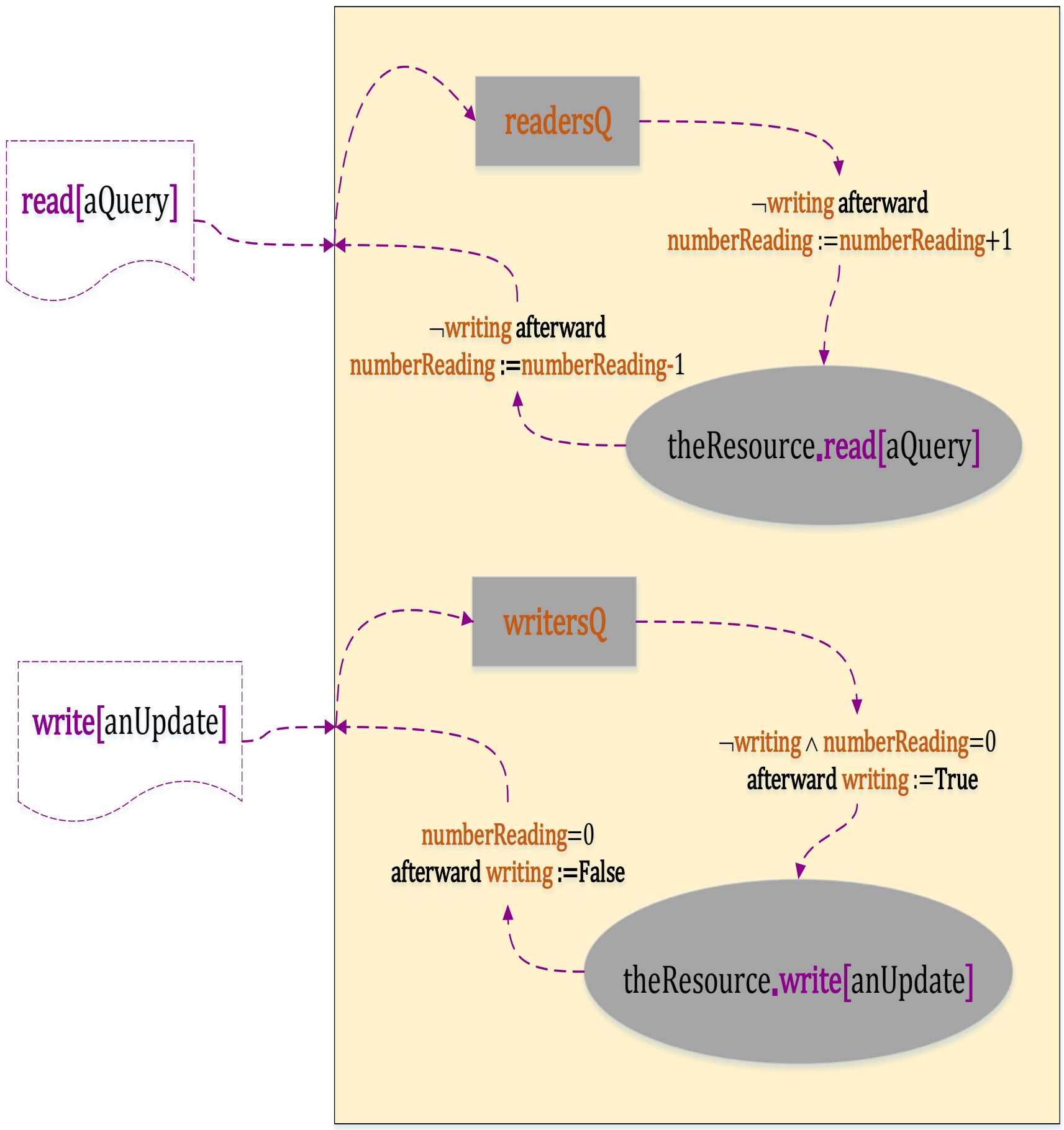

Note:

1. At most one activity is allowed to execute in the cheese.[i]
2. The cheese has holes.[ii]
3. The value of a variable[iii] cannot change in the cheese.[iv]

---

[i] Cheese is yellow in the diagram
[ii] A hole is grey in the diagram
[iii] A variable is orange in the diagram
[iv] Of course, other external Actors can change.

**Inconsistency Robustness:**
**Clausal Concurrent Programming Languages**

The combination of efficient inheritance-based virtual procedure invocation (pioneered in Simula) together with class libraries and browsers (pioneered in Smalltalk) provided better tools than the slower pattern-directed invocation of the FGCS Prolog-style clause programs. Consequently, the ICOT program languages never took off and instead object-oriented program languages like Java and JavaScript became the mainstream.

***Downfall***

The technical managers at ICOT were aware of some of the pitfalls that had tripped up previous Artificial Intelligence (AI) researchers. So they deliberately avoided calling ICOT an AI Project. Instead they had the vision of an integrated hardware/software system [Uchida and Fuchi 1992]. However, the Prolog-style clause-program paradigm turned not to be a suitable foundation because of poor modularity and lack of efficiency by comparison with direct message passing [Hewitt and Agha 1988]. Another problem was that multi-processors found it difficult to compete in the marketplace because at the time single processors were rapidly increasing in speed and connections between multiple processors suffered long latencies.

**The reliance on Prolog-style clause procedures was a principle contributing cause to the failure of ICOT to achieve commercial success.**

**Inconsistency Robustness: ICOT**

MITI's Fifth Generation strategy backfired because Japanese companies refused to productize ICOT hardware.

However, the architects of ICOT did get some things right:

- The project largely avoided the Mental Agent paradigm [Hewitt 2009]
- The project correctly placed tremendous emphasis on research in concurrency and parallelism as an emerging computing paradigm.

## What is a Logic Program?

> "*It would be like saying Prolog and SLD-Resolution is the only way to do Logic Programming.* ***To some extent, the LP*** ["Logic Programming"] ***community's insistence on clinging to this "exclusive method" has contributed to the relative disinterest in LP following its development in the 1980's and 1990's.*** " [Aït-Kaci 2009] (Emphasis added.)

Developments by different research groups in the fall of 1972 gave rise to a controversy over Logic Programs that persists to this day in the form of following alternatives:

1. Logic Programs using *procedural interpretation of logic-clause syntax for procedures* [Kowalski 2014][i]
2. Logic Programs in which *each computational step*[ii] *is logically inferred* [iii]

This article argues for the second conception based on the following considerations:

- Logic-clause syntax is inadequate to express Logic Programs, e.g., logic-clause syntax lacks the ability to construct sets using {x |⊩ Φ[x]}, where Φ is a grounded-complete predicated. For example, the following Logic Program is not adequately expressible using logic-clause syntax:[iv]
  **When** ⊩ Distance[*start, finish, aDistance*]→
  ⊢ *aDistance*=Minimum.[{*nextDistance* + *remainingDistance*
  | ⊩ Street[*aName, start, next, nextDistance*],
  Distance[*next, finish, remainingDistance*]}[37]]▮
- Programs in logic-clause syntax are a special case of the second alternative because each computational step of a program in logic-clause syntax is logically inferred by forward chaining and backward chaining.
- Reducing propositions to logic-clause syntax can obscure their natural structure

---

[i] Logic-clause syntax for a logic program is $\Psi \Leftarrow (\Phi_1 \wedge \ldots \wedge \Phi_n)$ where Ψ and each of the $\Phi_i$ is either $P[t_1, \ldots, t_m]$ or $\neg P[t_1, \ldots, t_m]$ for some atomic predicate P and terms $t_j$.

[ii] in the Actor Model of computation

[iii] *e.g.* using Direct Logic

[iv] See discussion earlier in this article of the program below

- Procedural interpretation of logic-clause syntax for procedures can obscure the natural structure of proofs.[i]

Logic Programs need to deal with both of the following:

- *Mathematical Theories* that are thought to be consistent by the overwhelming consensus of professional working mathematicians, *e.g.* Hilbert spaces, homology theory, *etc*.
- *Theories of Practice* that are pervasively inconsistent, *e.g.,* theories of climate change and the human brain. Mathematical theories are often freely used *within* theories of practice.

Going beyond the limitations of logic-clause syntax, there is a core subset of Logic Program constructs that are applicable to both Classical Direct Logic and Inconsistency Robust Direct Logic.[ii]

**Inconsistency Robustness: Characterization of Logic Programs**

Developments by different research groups in the fall of 1972 gave rise to a controversy over Logic Programs that persists to this day in the form of following alternatives:

1. Each computational step (*e.g.* as defined in the Actor Model) of a Logic Program is logically inferred (*e.g.* in Direct Logic) with explicit assertions and goals processed using pattern-directed invocation.
2. A Logic Program is expressed using logic-clause syntax that is interpreted as both a backward-chaining program and a forward-chaining program.

[i] *e.g.* ActorScript Logic Programs using Inconsistency Robust Natural Deduction in Direct Logic

[ii] See appendix of this article.

## Logic Programs versus Kowalski's "Logic Programming"

Kowalski has for a long time advocated "*Logic Programming*" characterized as follows:[38]

> *The driving force behind logic programming is the idea that a single formalism suffices for both logic and computation, and that logic subsumes computation. ...*
>
> *Logic programming aims ... to unify different areas of computing by exploiting the greater generality of logic* [over computational models[i]]. *It does so by building upon and extending*[ii] *one of the simplest, yet most powerful logics imaginable, namely the logic of Horn clauses.*

Kowalski's Logic Programming is a scientific research programme in the sense of [Lakatos 1980]. According to [Kowalski 2006]:

> *Admittedly, I have been messianic in my advocacy of Logic, and I make no apologies for it. Pushing Logic as hard as I could has been my way of trying to discover its limits.*

Kowalski's version of Logic Programing has the following limitations:

- based on the following mistaken assumptions:
  - "*that logic subsumes computation*"[iii]
  - that Horn clauses are the basis for a powerful logic[iv]
- lacks inconsistency-robust inference because it is based on classical logic making use of resolution theorem proving[v]
- lacks the precision of a well-defined scientific concept.[vi]

## Overview

[i] *e.g.*, Turing Machines, relational model of data base queries, the Actor Model, *etc.*

[ii] using logic-clause syntax for a Logic Program, *e.g.*, $\Psi \Leftarrow (\Phi_1 \wedge \ldots \wedge \Phi_n)$ where $\Psi$ and each of the $\Phi_i$ is either $P[t_1, \ldots, t_m]$ or $\neg P[t_1, \ldots, t_m]$ for some atomic predicate P and terms $t_j$.

[iii] Contrary to Kowalski, there are computations that cannot be performed using logical inference. See discussion earlier in this article.

[iv] Contrary to Kowalski, the logic-clause generalization of Horn-clause syntax is inadequate for expressing Logic Programs as explained earlier in this article. Also, Horn clauses do *not* form the basis for the most powerful logic because they are based on first-order logic, which has unintended models of axioms. For example, first-order logic cannot even characterize the integers up to isomorphism. See discussion in article 1-2 in this volume.

[v] As explained earlier in this article, resolution theorem provers can make invalid inferences using inconsistent axioms.

[vi] which could be fixed by defining the concept as "programming using Logic Programs" for a well-defined definition of a Logic Program.

| Contradiction | Outcome |
|---|---|
| Kowalski advocates using resolution theorem proving to make inferences about inconsistent information systems. Unfortunately, using resolution theorem provers allows inferring every proposition from an inconsistent information system. | Using Direct Logic, inconsistent-robust inference allows inference about inconsistent information systems without enabling inference of every sentence. |
| Planner was intended to be a general purpose programming language for the procedural embedding of knowledge. Partly as a result of limitations of contemporary computers, pragmatic decisions were made in the implementation of Planner that limited its generality. | Programming languages (like ActorScript) have been developed that incorporate general Logic Programs. |
| Kowalski, *et. al.* advocated that Horn-clause syntax for programs be used as the foundation of Logic Programs. But Horn-clause syntax for programs (and slightly more general logic-clause syntax) lack the generality and modularity needed for Logic Programs. | Programming languages (like ActorScript) are not restricted to logic-clause syntax for Logic Programs. |
| The Japanese Fifth Generation Project (ICOT) attempted to create a new computer architecture based on logic-clause programs. | ICOT failed to gain commercial traction although it was an important research project. |
| Kowalski claims that mathematical logical deduction subsumes computation. However, there are computations that cannot be implemented using deduction (*i.e.* Logic Programs) and there are important applications that cannot be implemented using only Logic Programs. | Actor programming languages (like ActorScript) use direct message passing to Actors and are not restricted to just Logic Programs. |
| Previously, change was modeled in Logic Programs using a global time that is physically meaningless. | The Actor model enables change in computation to be modeled by causal partial orders of message-passing events. |

The above examples are intended to be case studies in Inconsistency Robustness in which information is formalized, contradictions are derived using Inconsistency Robust reasoning, and arguments are formalized for and against contradictory propositions. A challenge for the future is to automate the reasoning involved in these case studies.

## Conclusion

> *Max Planck, surveying his own career in his Scientific Autobiography [Planck 1949], sadly remarked that 'a new scientific truth does not triumph by convincing its opponents and making them see the light, but rather because its opponents eventually die, and a new generation grows up that is familiar with it.'*
> [Kuhn 1962]

A fundamental principle of Inconsistency Robustness is to make contradictions explicit so that arguments for and against propositions can be formalized. This paper has explored the role of Inconsistency Robustness in the history and theory of Logic Programs by making contradictions explicit and using them to explicate arguments. The development of Logic Programs has been shown to be a productive area for applying principles of Inconsistency Robustness.

The examples presented in this paper are intended to be case studies in Inconsistency Robustness in which information is formalized, contradictions are derived using Inconsistency Robust reasoning, and arguments are formalized for and against contradictory propositions with developments that included:

- Some arguments were dropped.
- Some arguments triumphed.
- Arguments were combined and improved.
- New arguments were developed

A challenge for the future is to automate the reasoning involved in these case studies.[i]

---

[i] Computerization of argumentation is still in its infancy, *cf.*, [Toulmin 1959, Woods 2000, Bench-Capon 2012]. Also, there is a great deal of ongoing research in formalizing mathematical proofs [Avigad and Harrison 2014].

**Acknowledgements**
Scott Fahlman made suggestions and comments that materially improved an earlier version of this article. Richard Waldinger made an important suggestion for improving the definition of Logic Programs and otherwise greatly helped to improve the paper. Jeremy Forth made helpful suggestions. Bob Kowalski and Pat Hayes made important contributions and suggestions that greatly improved an earlier version article. I especially thank Bob for sharing his view of his recent research. Maarten van Emden corrected some typos. Dale Schumacher provided useful suggestions in the section on ICOT.

In 2013, this article was substantially revised to address issues of Inconsistency Robustness. Bob Kowalski provided useful comments on the abstract, which contributed to better delineation similarities and differences in our views. Alan Bundy pointed out many crucial places where the presentation needed improvement. Alan Karp pointed out a major bug in the Logic Program in the end notes. Ken Kahn provided helpful comments on clausal concurrent programming languages. Extensive discussions with Eric Kao helped clarify the discussion of the work of Kowalski. Harold Boley provided helpful suggestions on the presentation.

Of course, any remaining errors are entirely my own.

## Bibliography

Hal Abelson and Gerry Sussman *Structure and Interpretation of Computer Programs (2nd edition)* MIT Press. 1996.

Hassan Aït-Kaci. *Children's Magic Won't Deliver the Semantic Web* CACM. March 2009

Robert Anderson and Woody Bledsoe (1970) *A Linear Format for Resolution with Merging and a New Technique for Establishing Completeness* JACM 17.

Bruce Anderson. *Documentation for LIB PICO-PLANNER* School of Artificial Intelligence, Edinburgh University. 1972.

Bruce Anderson and Pat Hayes. *The logician's folly* DCL Memo 54. University of Edinburgh. 1972.

Jeremy Avigad and John Harrison. *Formally Verified Mathematics* CACM. April 2014.

Bruce Baumgart. *Micro-Planner Alternate Reference Manual Stanford* AI Lab Operating Note No. 67, April 1972.

L. Bertossi and J. Chomicki. *Query answering in inconsistent databases* Logics for Emerging Applications of Databases. 2003.

Trevor Bench-Capon. *The Long and Winding Road: Forty Years of Argumentation* COMMA'12.

Philippe Besnard and Anthony Hunter. *Quasi-classical logic: Non-trivializable classical reasoning from inconsistent information* Symbolic and Quantitative Approaches to Uncertainty. Springer LNCS 1995.

Maurice Bruynooghe, Luís Pereira, Jörg Siekmann, and Maarten van Emden. *A Portrait of a Scientist as a Computational Logician Computational Logic: Logic Programming and Beyond: Essays in Honour of Robert A. Kowalski, Part I* Springer. 2004.

Andrea Cantini, *Paradoxes and Contemporary Logic*, Stanford Encyclopedia of Philosophy. Winter 2012 Edition.

Eugene Charniak *Toward a Model of Children's Story Comprehension* MIT AI TR-266. December 1972.

Alonzo Church *A Set of postulates for the foundation of logic* Annals of Mathematics. Vol. 33, 1932. Vol. 34, 1933.

Paolo Ciancarini and Giorgio Levi. *What is Logic Programming good* for in Software Engineering? International Conference on Software Engineering. 1993.

Will Clinger. *Foundations of Actor Semantics* MIT Mathematics Doctoral Dissertation. June 1981

Jacques Cohen. *A view of the origins and development of Prolog* CACM. January 1988.

Alain Colmerauer and Philippe Roussel. *The birth of Prolog* History of Programming Languages. ACM Press. 1996.

Ole-Johan Dahl and Kristen Nygaard. *Class and subclass declarations* IFIP TC2 Conference on Simulation Programming Languages. May 1967.

John Dawson. *Shaken Foundations or Groundbreaking Realignment?* 2006.

Julian Davies. *Popler 1.6 Reference Manual* University of Edinburgh, TPU Report No. 1, May 1973.

Julian Davies. *Representing Negation in a Planner System* AISB'74.

Adriaan De Groot *Thought and choice in Chess* Mouton De Gruyter. 1965.

Jason Eisner and Nathaniel W. Filardo. *Dyna: Extending Datalog for modern AI.* Datalog Reloaded. Springer. 2011.

Hendrik Decker. *Historical and Computational Aspects of Paraconsistency in View of the Logic Foundation of Databases.* Semantics in Databases. Springer. 2003.

Hendrik Decker. *A Case for Paraconsistent Logic as a Foundation of Future Information Systems.* CAiSE'05 Workshop of PHISE'05. 2005.

Hendrik Decker. *How to Confine Inconsistency or, Wittgenstein only Scratched the Surface.* ECAP10. October 4-6, 2010.

Maarten van Emden. *The Early Days of Logic Programming: A Personal Perspective Association of Logic Programming* Newsletter. August 2006.

Scott Fahlman. *A Planning System for Robot Construction Tasks* MIT AI TR-283. June 1973.

Solomon Feferman *Axioms for determinateness and truth* Review of Symbolic Logic. 2008.

J.M. Foster and E.W. Elcock. *ABSYS: An Incremental Compiler for Assertions* Machine Intelligence 4. Edinburgh University Press. 1969.

Kazuhiro Fuchi, Robert Kowalski, Kazunori Ueda, Ken Kahn, Takashi Chikayama, and Evan Tick. *Launching the new era* CACM. 1993

Mike Gordon, Robin Milner, and Christopher Wadsworth. *Edinburgh LCF: A Mechanized Logic of Computation* Springer-Verlag. 1979.

Cordell Green. *Application of Theorem Proving to Problem Solving* IJCAI'69.

Steve Gregory. *Concurrent Logic Programming Before ICOT: A Personal Perspective* August 15, 2007.

Irene Greif. *Semantics of Communicating Parallel Professes* MIT EECS Doctoral Dissertation. August 1975.

Pat Hayes *Computation and Deduction* Mathematical Foundations of Computer Science: Proceedings of Symposium and Summer School, Štrbské Pleso, High Tatras, Czechoslovakia, September 3-8, 1973.

Pat Hayes. *Semantic Trees* Ph.D. thesis. Edinburgh University. 1973.

Pat Hayes. *Some Problems and Non-Problems in Representation Theory* AISB'74.

Carl Hewitt *PLANNER: A Language for Proving Theorems in Robots* IJCAI'69.

Carl Hewitt *Procedural Embedding of Knowledge in Planner* IJCAI 1971.

Carl Hewitt *Description and Theoretical Analysis (Using Schemata) of Planner, A Language for Proving Theorems and Manipulating Models in a Robot* AI Memo No. 251, MIT Project MAC, April 1972.

Carl Hewitt, Peter Bishop and Richard Steiger. *A Universal Modular Actor Formalism for Artificial Intelligence* IJCAI'73.

Carl Hewitt *Stereotypes as an Actor Approach Towards Solving the Problem of Procedural Attachment in Frame Theories* Theoretical Issues In Natural Language Processing. 1975.

Carl Hewitt. Viewing Control Structures as Patterns of Passing Messages AI Memo 410. December 1976. Journal of Artificial Intelligence. June 1977.

Carl Hewitt and Henry Baker *Actors and Continuous Functionals* Proceeding of IFIP Working Conference on Formal Description of Programming Concepts. August 1-5, 1977.

Carl Hewitt *The Challenge of Open Systems* Byte Magazine. April 1985.

Carl Hewitt and Jeff Inman. *DAI Betwixt and Between: From 'Intelligent Agents' to Open Systems Science* IEEE Transactions on Systems, Man, and Cybernetics. Nov. /Dec. 1991.

Carl Hewitt and Gul Agha. *Guarded Horn clause languages: are they deductive and Logical?* International Conference on Fifth Generation Computer Systems, Ohmsha 1988. Tokyo. Also in Artificial Intelligence at MIT, Vol. 2. MIT Press 1991

Carl Hewitt (2006b) *What is Commitment? Physical, Organizational, and Social* COIN@AAMAS'06.

Carl Hewitt (2008a) *A historical perspective on developing foundations iInfo™ information systems: iConsult™ and iEntertain™ apps using iDescribers™ information integration for iOrgs™ information systems* (revised version of *Development of Logic Programming: What went wrong, What was done about it, and What it might mean for the future* Proceedings of What Went Wrong and Why: Lessons from AI Research and Applications edited by Mehmet Göker and Daniel Shapiro. AAAI Press. AAAI'08.) Google Knol.

Carl Hewitt (2013) *ActorScript(TM) extension of C sharp (TM), Java(TM), and Objective C(TM): iAdaptive(TM) concurrency for antiCloud(TM) privacy and security* arXiv:1008.2748

Carl Hewitt (2009) *Perfect Disruption: Causing the Paradigm Shift from Mental Agents to ORGs* IEEE Internet Computing. Jan/Feb 2009.

Carl Hewitt (2010) *Actor Model of Computation* arXiv:1008.1459

Carl Hewitt (2011), *Formalizing common sense for inconsistency-robust Information Integration using Direct Logic™ and the Actor Model* ArXiv 0907.3330

Carl Hewitt. *Health Information Systems Technologies.* Stanford CS Colloquium. June 2012. http://HIST.carlhewitt.info

Carl Hewitt (2013). *Inconsistency Robustness in Foundations: Mathematics self proves its own Consistency and Other Matters* to appear in Inconsistency Robustness 2014.

Alfred Horn. *On sentences which are true of direct unions of algebras* Journal of Symbolic Logic. March 1951.

Matthew Huntbach and Graem Ringwood. *Agent-Oriented Programming: From Prolog to Guarded Definite Clauses* Springer. 1999.

Anthony Hunter. *Reasoning with Contradictory Information using Quasi-classical Logic* Journal of Logic and Computation. Vol. 10 No. 5. 2000.

Daniel Ingalls. *The Evolution of the Smalltalk Virtual Machine* Smalltalk-80: Bits of History, Words of Advice. Addison Wesley. 1983.

Stanisław Jaśkowski *On the Rules of Suppositions in Formal Logic* Studia Logica 1, 1934. (reprinted in: *Polish logic 1920-1939*, Oxford University Press, 1967.

Ken Kahn and Vijay Saraswat. *Actors as a Special Case of Concurrent Constraint Programming* ECOOP/OOPSLA. October 1990.

Rajesh Karmani and Gul Agha. *Actors*. Encyclopedia of Parallel Computing 2011.

Michael Kassoff, Lee-Ming Zen, Ankit Garg, and Michael Genesereth. *PrediCalc: A Logical Spreadsheet Management System* 31st International Conference on Very Large Databases (VLDB). 2005.

Stephen Kleene and John Barkley Rosser *The inconsistency of certain formal logics* Annals of Mathematics Vol. 36. 1935.

William Kornfeld and Carl Hewitt T*he Scientific Community Metaphor* MIT AI Memo 641. January 1981.

William Kornfeld (1981a) *The Use of Parallelism to Implement a Heuristic Search* IJCAI'81.
William Kornfeld (1981b) *Parallelism in Problem Solving* MIT EECS Doctoral Dissertation. August 1981.
William Kornfeld. *Combinatorially Implosive Algorithms* CACM. 1982.
Robert Kowalski and Pat Hayes. *Semantic trees in automatic theorem-proving* Machine Intelligence 4. Edinburgh Press. 1969.
Robert Kowalski *Predicate Logic as Programming Language* IFIP'74.
Robert Kowalski *Logic for problem-solving* DCL Memo 75. Dept. of Artificial Intelligence. Edinburgh. 1974.
Robert Kowalski. *A proof procedure using connection graphs* JACM. October 1975.
Robert Kowalski *Algorithm = Logic + Control* CACM. July 1979.
Robert Kowalski. *Response to questionnaire* Special Issue on Knowledge Representation. SIGART Newsletter. February 1980.
Robert Kowalski *The Limitations of Logic* Proceedings of the ACM Annual Conference on Computer Science. 1986.
Robert Kowalski (1988a) *The Early Years of Logic Programming* CACM. January 1988.
Robert Kowalski (1988b) *Logic-based Open Systems Representation and Reasoning.* Stuttgart Conference Workshop on Discourse Representation, Dialogue tableaux and Logic Programming. 1988.
Robert Kowalski. *Logic Programming* MIT Encyclopedia of Cognitive Science. MIT Press. 1999.
Robert Kowalski. *Logic Programming and the Real World* Logic Programming Newsletter. January 2001.
Robert Kowalski (2004a) *History of the Association of Logic Programming* October 2004.
Robert Kowalski (2004b) *Directions for Logic Programming* Computational Logic: Logic Programming and Beyond: Essays in Honour of Robert A. Kowalski, Part I Springer. 2004.
Robert Kowalski *Re: Which Logicists?* Email to Carl Hewitt, Pat Hayes, Michael Genesereth, Richard Waldinger, and Mike Dunn. December 20, 2006.
Robert Kowalski (2007b) *Philosophy* Wikipedia. April 17, 2007.
Robert Kowalski (2007b) Robert Kowalski (2007b) *Philosophy* Wikipedia. January 10, 2007.
Robert Kowalski *Reasoning with Conditionals in Artificial Intelligence* The Psychology of Conditionals Oxford University Press. 2008.
Robert Kowalski *Email to Carl Hewitt* April 2009.
Robert Kowalski *Computational Logic and Human Thinking: How to be Artificially Intelligent* Cambridge University Press. 2011.
Robert Kowalski. *History of Logic Programming* Computational Logic Volume 9. Elsevier. 2014.
Robert Kowalski and Fariba Sadri. *Reactive Computing as Model Generation*

New Generation Computing. 2015.
Thomas Kuhn. *The Structure of Scientific Revolutions* University of Chicago Press. 1962.
Imre Lakatos. *The Methodology of Scientific Research Programmes: Volume 1: Philosophical Papers*. Cambridge University Press. 1980.
Imre Lakatos. *Proofs and Refutations* Cambridge University Press. 1976.
Peter Landin *A Generalization of Jumps and Labels* Report UNIVAC Systems Programming Research. August 1965. Reprinted in Higher Order and Symbolic Computation. 1998.
Bruno Latour *Science in Action: How to Follow Scientists and Engineers through Society* Harvard University Press. 1988.
Clarence Lewis and Cooper Langford. *Symbolic Logic* Century-Croft, 1932.
James Lighthill *Artificial Intelligence: A General Survey* Artificial Intelligence: a paper symposium. UK Science Research Council. 1973.
Donald MacKenzie *Mechanizing Proof* MIT Press. 2001.
E. Mayol and E. Teniente. *A survey of current methods for integrity constraint maintenance and view updating* ER Workshops. 1999.
John McCarthy *Programs with common sense* Symposium on Mechanization of Thought Processes. National Physical Laboratory, UK. Teddington, England. 1958.
John McCarthy, Paul Abrahams, Daniel Edwards, Timothy Hart, and Michael Levin. *Lisp 1.5 Programmer's Manual* MIT Computation Center and Research Laboratory of Electronics. 1962.
John McCarthy *Review of 'Artificial Intelligence: A General* Survey' Artificial Intelligence: a paper symposium. UK Science Research Council. 1973.
John McCarthy. *Sterile Containers*
www.ai.sri.com/~rkf/designdoc/sterile.ps September 8, 2000.
L. Thorne McCarty. *Reflections on TAXMAN: An Experiment on Artificial Intelligence and Legal Reasoning* Harvard Law Review. Vol. 90, No. 5, March 1977.
Drew McDermott and Gerry Sussman *The Conniver Reference Manual* MIT AI Memo 259A. January 1974.
Drew McDermott *The Prolog Phenomenon* ACM SIGART Bulletin. Issue 72. July, 1980.
Robin Milner. *Logic for Computable Functions: description of a machine implementation.* Stanford AI Memo 169. May 1972
Marvin Minsky (ed.) *Semantic Information Processing* MIT Press. 1968.
Marvin Minsky and Seymour Paper. *Progress Report, Artificial Intelligence* MIT AI Memo 252. 1972.
Alexander Nekham. *De Naturis Rerum* 1200. republished in Thomas Wright, editor. London: Longman, 1863.
Allen Newell and Herbert Simon *The logic theory machine: A complex information processing system* IRE Trans. Information Theory IT-2:61-79. 1956.

Nils Nilsson *Artificial Intelligence: A New Synthesis* San Francisco: Morgan Kaufmann, 1998.

L. Orman. *Transaction repair for integrity enforcement* TKDE, 13(6). 2001.

Mike Paterson and Carl Hewitt. *Comparative Schematology* MIT AI Memo 201. August 1970.

Max Planck *Scientific Autobiography and Other Papers* 1949.

George Polya (1957) *Mathematical Discovery: On Understanding, Learning and Teaching Problem Solving Combined Edition* Wiley. 1981.

Karl Popper (1935, 1963) *Conjectures and Refutations: The Growth of Scientific Knowledge Routledge*. 2002.

John Alan Robinson *A Machine-Oriented Logic Based on the Resolution Principle.* CACM. 1965.

Kenneth Ross, Yehoshua Sagiv. *Monotonic aggregation in deductive databases.* Principles of Distributed Systems. June 1992.

Jeff Rulifson, Jan Derksen, and Richard Waldinger *QA4, A Procedural Calculus for Intuitive Reasoning* SRI AI Center Technical Note 73, November 1973.

Philippe Rouchy. *Aspects of PROLOG History: Logic Programming and Professional Dynamics* TeamEthno-Online Issue 2. June 2006.

Earl Sacerdoti, et al., *QLISP A Language for the Interactive Development of Complex Systems* AFIPS. 1976.

Erik Sandewall. *From Systems to Logic in the Early Development of Nonmonotonic Reasoning CAISOR. July, 2006.*

Gary Saxonhouse *What's All This about Japanese Technology Policy?* Cato Institute. August 17, 2001

Dana Scott *Data Types as Lattices.* SIAM Journal on computing. 1976.

Marek Sergot. *Bob Kowalski: A Portrait* Computational Logic: Logic Programming and Beyond: Essays in Honour of Robert A. Kowalski, Part I Springer. 2004.

Jon Sigurdson *Industry and state partnership: The historical role of the engineering research associations in Japan* 1986

Ehud Shapiro *The family of concurrent logic programming languages* ACM Computing Surveys. September 1989.

Gerry Sussman, Terry Winograd and Eugene Charniak *Micro-Planner Reference Manual (Update)* AI Memo 203A, MIT AI Lab, December 1971.

Gerry Sussman and Guy *Steele Scheme: An Interpreter for Extended Lambda Calculus* MIT AI Lab Memo 349. December 1975.

Shunichi Uchida and Kazuhiro *Fuchi Proceedings of the FGCS Project Evaluation Worksho*p Institute for New Generation Computer Technology (ICOT). 1992.

Martin van Emden and Robert Kowalski. *The semantics of predicate logic as a programming language* Edinburgh TR 1974. JACM'76.

Arthur Prior. *The runabout inference ticket* Analysis, 21, 1960-61.

Alfred Tarski and Robert Vaught (1957). *Arithmetical extensions of relational systems* Compositio Mathematica 13.

Terry Winograd *Procedures as a Representation for Data in a Computer Program for Understanding Natural Language* MIT AI TR-235. January 1971.

John Woods. *How Philosophical is Informal Logic?* Informal Logic. 20(2). 2000

Larry Wos, *et al. Efficiency and completeness of the set of support strategy in theorem proving.* JACM), 12(4), 1965.

Stephen Toulmin. *The uses of argument*. 1959.

## Appendix. Inconsistency Robust Logic Programs

### Notation of Direct Logic

> The aims of logic should be the creation of "*a unified conceptual apparatus which would supply a common basis for the whole of human knowledge.*" [Tarski 1940]

In Direct Logic, unrestricted recursion is allowed in programs. For example,

- There are uncountably many Actors.[39] For example, Real▪[ ] can output any real number[i] between 0 and 1 where

  Real▪[ ] ≡ [(0 **either** 1), ⩛**Postpone** Real▪[ ]]

  where

  - (0 **either** 1) is the nondeterministic choice of 0 or 1,
  - [*first*, ⩛*rest*] is the list that begins with *first* and whose remainder is *rest*, and

- **Postpone** *expression* delays execution of *expression* until the value is needed.

- There are uncountably many propositions (because there is a different proposition for every real number). Consequently, there are propositions that are not the abstraction of any element of a denumerable set of sentences. For example,

  $p \equiv [x\in\mathbb{R}]\rightarrow([y\in\mathbb{R}]\rightarrow(y=x))$

  defines a different predicate p[x] for each real number x, which holds for only one real number, namely x.[ii]

It is important to distinguish between strings, sentences, and propositions. Some strings can be parsed into sentences[iii], which can be abstracted into propositions that can be asserted. Furthermore, grammar terms[iv] can be abstracted into Actors (*i.e.* objects in mathematics).

---

[i] using binary representation.

[ii] For example (p[3])[y] holds if and only if y=3.

[iii] which are grammar tree structures

[iv] which are grammar tree structures

Abstraction and parsing are becoming increasingly important in software engineering. *e.g.,*

- The execution of code can be dynamically checked against its documentation. Also Web Services can be dynamically searched for and invoked on the basis of their documentation.
- Use cases can be inferred by specialization of documentation and from code by automatic test generators and by model checking.
- Code can be generated by inference from documentation and by generalization from use cases.

**Abstraction and parsing are needed for large software systems so that that documentation, use cases, and code can mutually speak about what has been said and their relationships.**

For example:

### Proposition

*e.g.* ∀[n:ℕ]→ ∃[m:ℕ]→ m>n
*i.e., for every* ℕ *there is a larger* ℕ

### Sentence

*e.g.* ⌈"∀[n:ℕ]→ ∃[m:ℕ]→ m>n"⌉

*i.e., the sentence that for every* ℕ *there is a larger* ℕ

### String

*e.g.* "∀[n:ℕ]→ ∃[m:ℕ]→ m>n"

*which is a string that begins with the symbol* "∀"

In Direct Logic, a sentence is a grammar tree (analogous to the ones used by linguists). Such a grammar tree has terminals that can be constants. And there are uncountably many constants, *e.g.*, the real numbers:

> The sentence ⌈3.14159... < 3.14159... + 1⌉ is impossible to obtain by parsing a string (where 3.14159... is an Actor[i] for the transcendental real number that is the ratio of a circle's circumference to its diameter). The issue is that there is no string which when parsed is ⌈3.14159... < 3.14159... + 1⌉
>
> Of course, because the digits of 3.14159... are computable, there is a $\text{term}_1$ such that ⌊$\text{term}_1$⌋ = 3.14159... that can be used to create the sentence ⌈$\text{term}_1$ < $\text{term}_1$ + 1⌉.
>
> However the sentence ⌈$\text{term}_1$ < $\text{term}_1$ + 1⌉ is not the same as ⌈3.14159... < 3.14159... + 1⌉ because it does not have the same vocabulary and it is a much larger sentence that has many terminals whereas ⌈3.14159... < 3.14159... + 1⌉ has just 3 terminals:

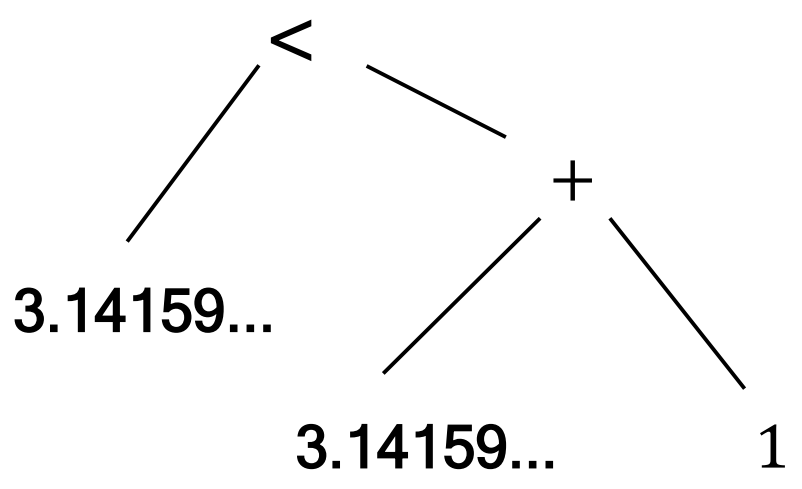

Consequently, sentences *cannot* be enumerated and there are some sentences that *cannot* be obtained by parsing strings. These arrangements exclude known paradoxes from Classical Direct Logic.[ii]

**Note: type theory of Classical Direct Logic is much stronger than constructive type theory with constructive logic[40] because Classical Direct Logic has all of the power of Classical Mathematics.**

---

[i] whose digits are incrementally computable

[ii] Please see historical appendix of this article.

Types and Propositions are defined as follows:

- ***Types***
  - **Boolean**, ℕ[41], **Sentence**, **Proposition**, **Proof**, **Type**, **Theory**:**Type**.
  - If $\sigma_1, \sigma_2$:**Type**, then $\sigma_1 \sqcup \sigma_2$, $[\sigma_1, \sigma_2]$[42], $[\sigma_1] \mapsto \sigma_2$[43], ${\sigma_2}^{\sigma_1}$[44]:**Type**.
  - If $\sigma$:**Type**, then **Term**$\triangleleft\sigma\triangleright$[45]:**Type**.
  - If $\sigma_1, \sigma_2$:**Type**, $f$:${\sigma_2}^{\sigma_1}$ and $x$:$\sigma_1$, then $f[x]$:$\sigma_2$.
  - If $\sigma_1, \sigma_2$:**Type**, then $\sigma_1 \sqcup \sigma_2$, $[\sigma_1] \mapsto \sigma_2$, ${\sigma_2}^{\sigma_1}$:**Type**
  - If $\sigma$:**Type**, then **Term**$\triangleleft\sigma\triangleright$:**Type**
- ***Propositions***, *i.e.*, x:**Proposition** ⇔ x constructed by the rules below:
  - If $\sigma$:**Type**, $\Pi$:**Boolean**$^{\sigma}$ and $x$:$\sigma$, then $\Pi[x]$:**Proposition**.[i]
  - If $\Phi$:**Proposition**, then $\neg\Phi$:**Proposition**.
  - If $\Phi$,$\Psi$:**Proposition**, then $\Phi \wedge \Psi$, $\Phi \vee \Psi$, $\Phi \Rightarrow \Psi$, $\Phi \Leftrightarrow \Psi$:**Proposition**.
  - If $p$:**Boolean** and $\Phi$,$\Psi$:**Proposition**, then
    (p � **True**⸰ $\Phi_1$, **False**⸰ $\Phi_2$):**Proposition**.[46]
  - If $\sigma_1$,$\sigma_2$:**Type**, $x_1$:$\sigma_1$ and $x_2$:$\sigma_2$, then
    $x_1 = x_2, x_1 \in x_2, x_1 \sqsubseteq x_2, x_1 \subseteq x_2, x_1 : x_2$:**Proposition**.
  - If $T$:**Theory** and $\Phi_{1\ \text{to}\ n}$:**Proposition**, then
    $(\Phi_1, \ldots, \Phi_k \vdash_T \Phi_{k+1}, \ldots, \Phi_n)$:**Proposition** [47]
  - If $T$:**Theory**, $p$:**Proof** and $\Phi$:**Proposition**, then
    $(\vdash^{p}_{T} \Phi)$:**Proposition**[48]

---

[i] $\Pi[x] \Leftrightarrow (\Pi[x] = \text{True})$

Note that $\sigma$:**Type**, $\Pi$:**Boolean**$^{\sigma}$ means that there are no fixed points for propositions.

Grammar trees (*i.e.* expressions, terms, and sentences) are defined as follows:

- ***Expressions**, i.e.,* x:**Expression**◁σ▷ ⇔ x constructed by the rules below:
  - **True**, **False**:**Constant**◁**Boolean**▷ and **0**,**1**:**Constant**◁ℕ▷.
  - If **σ**:**Type** and **x**:**Constant**◁**σ**▷, then **x**:**Expression**◁**σ**▷.
  - If **σ**:**Type** and **x**:**Variable**◁**σ**▷, then **x**:**Expression**◁**σ**▷.
  - If **σ**,$\sigma_{1 \text{ to } n}$:**Type**, $x_{1 \text{ to } n}$:**Expression**◁$\sigma_{1 \text{ to } n}$▷ and **y**:**Expression**◁**σ**▷, then (**Let** {$v_1 \equiv x_1, \ldots, v_n \equiv x_n$}, **y**):**Expression**◁**σ**▷ and $v_{1 \text{ to } n}$:**Variable**◁$\sigma_{1 \text{ to } n}$▷ in **y** and in each $x_{1 \text{ to } n}$.
  - If $e_1$, $e_2$:**Expression**◁**Type**▷, then ⌈$e_1 \sqcup e_2$⌉,⌈[$e_1$, $e_2$]⌉,⌈[$e_1$]↦$e_2$⌉,⌈$e_2{}^{e_1}$⌉:**Expression**◁**Type**▷.
  - If **σ**:**Type**, $t_1$:**Expression**◁**Boolean**▷, $t_2$, $t_3$:**Expression**◁**σ**▷, then ⌈$t_1$ � **True** ⦂ $t_2$, **False** ⦂ $t_3$⌉:**Expression**◁**σ**▷.[49]
  - If $\sigma_1$,$\sigma_2$:**Type**, **t**:**Expression**◁$\sigma_2$▷, then ⌈[**x**:$\sigma_1$]→ **t**⌉:**Expression**◁[$\sigma_1$]↦$\sigma_2$▷ and **x**:**Variable**◁$\sigma_1$▷.[50]
  - If $\sigma_1$,$\sigma_2$:**Type**, **p**:**Expression**◁[$\sigma_1$]↦$\sigma_2$▷ and **x**:**Expression**◁$\sigma_1$▷, then ⌈**p**.[**x**] ⌉:**Expression**◁$\sigma_2$▷.
  - If **σ**:**Type** and **e**:**Expression**◁**σ**▷, then ⌈⌊**e**⌋⌉:**Expression**◁**σ**▷.
  - If **σ**:**Type** and **e**:**Expression**◁**σ**▷ with no free variables and **e** converges, then ⌊**e**⌋:**σ**.
- ***Terms**, i.e.,* x:**Term**◁σ▷ ⇔ x constructed by the rules below:
  - If **σ**:**Type** and **x**:**Constant**◁**σ**▷, then **x**:**Term**◁**σ**▷.
  - If **σ**:**Type** and **x**:**Variable**◁**σ**▷, then **x**:**Term**◁**σ**▷.
  - If $t_1$, $t_2$:**Term**◁**Type**▷, then ⌈$t_1 \sqcup t_2$⌉,⌈[$t_1$, $t_2$]⌉,⌈[$t_1$]↦$t_2$⌉,⌈$t_2{}^{t_1}$⌉:**Term**◁Type▷.
  - If **σ**:**Type**, $t_1$:**Term**◁**Boolean**▷, $t_2$,$t_3$:**Term**◁**σ**▷, then ⌈$t_1$ � **True** ⦂ $t_2$, **False** ⦂ $t_3$⌉:**Term**◁**σ**▷.
  - If $\sigma_1$,$\sigma_2$:**Type**, **f**:**Term**◁$\sigma_2{}^{\sigma_1}$▷ and **t**:**Term**◁$\sigma_1$▷, then ⌈**f**[**t**]⌉:**Term**◁$\sigma_2$▷.
  - If $\sigma_1$,$\sigma_2$:**Type** and **t**:**Term**◁$\sigma_2$▷, then ⌈[**x**:$\sigma_1$]→ **t**⌉:**Term**◁$\sigma_2{}^{\sigma_1}$▷ and **x**:**Variable**◁σ1▷ in **t**.
  - If **σ**:**Type** and **t**:**Term**◁**σ**▷, then ⌈⌊**t**⌋⌉:**Term**◁**σ**▷.
  - If **σ**:**Type**, **e**:**Expression**◁**σ**▷ with no free variables and **e** converges, then **e**:**Constant**◁**σ**▷.
  - If **σ**:**Type** and **t**:**Term**◁**σ**▷ with no free variables, then ⌊**t**⌋:**σ**.

- *Sentences, i.e.,* x:Sentence ⇔ x constructed by the rules below:
  - If $s_1$:Sentence then, ⌈¬$s_1$⌉:Sentence.
  - If $s_1$:Sentence and $s_2$:Sentence then ⌈$s_1$∧$s_2$⌉,⌈$s_1$∨$s_2$⌉,⌈$s_1$⇨$s_2$⌉,⌈$s_1$⇔$s_2$⌉:Sentence.
  - If σ:Type, $t_1$:Term◁Boolean$^{\sigma}$▷ and $t_2$:Term◁σ▷*,* then ⌈$t_1$[$t_2$]⌉:Sentence
  - If t:Term◁Boolean▷, $s_1$,$s_2$:Sentence, then ⌈t�True ⦂ $s_1$, False ⦂ $s_2$⌉:Sentence.[51]
  - If $\sigma_1$,$\sigma_2$:Type, $t_1$:Term◁$\sigma_1$▷ and $t_2$:Term◁$\sigma_2$▷*,* then ⌈$t_1$=$t_2$⌉,⌈$t_1$∈$t_2$⌉,⌈$t_1$⊑$t_2$⌉,⌈$t_1$⊆$t_2$⌉,⌈$t_1$:$t_2$⌉:Sentence.
  - If σ:Type and s:Sentence, then ⌈∀[x:σ]→ s⌉,⌈∃[x:σ]→ s⌉:Sentence and x:Variable◁σ▷ in s.
  - If T:Term◁Theory▷ and $s_{1\text{ to }n}$:Sentence*,* then ⌈$s_1$, …, $s_k$ $\vdash_T$ $s_{k+1}$, …, $s_n$⌉:Sentence
  - If T:Term◁Theory▷, p:Term◁Proof▷ and s:Sentence, then ⌈ $\vdash\frac{p}{T}$s⌉:Sentence
  - If s:Sentence, then ⌈⌊s⌋⌉:Sentence.
  - If s:Sentence with no free variables, then ⌊s⌋ :Proposition.

## Inconsistency Robust Logic Programs

In Logic Programs, computational steps are logically inferred.[i]

### *Forward Chaining*

Forward chaining is performed using ⊢

⟦"⊢"$_{Theory}$ *PropositionExpression* ⟧:*Expression*
Assert *PropositionExpression* for *Theory*.

⟦"**When**" "⊢"$_{Theory}$ *PropositionPattern* "→" *Expression*⟧:*Continuation*
When *PropositionPatterns* holds for *Theory*, evaluate *Expression*.

Illustration of forward chaining:
⊢$_t$ Human[Socrates]▮
**When** ⊢$_t$ Human[*x*]→ ⊢$_t$ Mortal[*x*]▮
will result in asserting Mortal[Socrates] for theory t

### *Backward Chaining*

Backward chaining is performed using ⊩

⟦"⊩"$_{Theory}$ *GoalPatterns* "→" *Expression*⟧:*Continuation*
Set *GoalPatterns* for *Theory* and when established evaluate *Expression*

⟦"⊩"$_{Theory}$ *GoalPattern*⟧:*Expression*
Set *GoalPattern* for *Theory* and return a list of assertions that satisfy the goal.

---

[i] [Church 1932; McCarthy 1963; Hewitt 1969, 1971, 2010; Milner 1972, Hayes 1973; Kowalski 1973].

(⟦"When" "⊩"$_{Theory}$ *GoalPattern* "→" *Expression*⟧):*Continuation*
When there are goals that matches *GoalPatterns* for *Theory*, evaluate *Expression*.

Illustration of backward chaining:

$\vdash_t$ Human[Socrates]▮

**When** $\Vdash_t$ Mortal[$x$]→ ($\Vdash_t$ Human[$x$]→ $\vdash_t$ Mortal[$x$])▮

$\Vdash_t$ Mortal[Socrates]▮

will result in asserting Mortal[Socrates] for theory t.

### *SubArguments*

This section explains how subarguments can be implemented in natural deduction.

**When** $\Vdash_s(\Psi \vdash_t \Phi)$→
 **Let** t' = Extension.[t],
  **Do** $\vdash_{t'} \Psi$,
   $\Vdash_{t'} \Phi$ → $\vdash_s(\Psi \vdash_t \Phi)$▮

Note that the following hold for t' because it is an extension of t:

- **When** $\vdash_t \Theta$→ $\vdash_{t'} \Theta$▮
- **When** $\Vdash_{t'} \Theta$→ $\Vdash_t \Theta$▮

### Inconsistency-Robust Propositional Equivalences

The following propositional equivalences hold in Inconsistency Robust Direct Logic:

- **Self Equivalence:** $\Psi \Leftrightarrow \Psi$
- **Double Negation:** $\neg\neg\Psi \Leftrightarrow \Psi$
- **Idempotence of $\wedge$:** $\Psi\wedge\Psi \Leftrightarrow \Psi$
- **Commutativity of $\wedge$:** $\Psi\wedge\Phi \Leftrightarrow \Phi\wedge\Psi$
- **Associativity of $\wedge$:** $\Psi \wedge (\Phi\wedge\Theta) \Leftrightarrow (\Psi\wedge\Phi) \wedge \Theta$
- **Distributivity of $\wedge$ over $\vee$:** $\Psi \wedge (\Phi\vee\Theta) \Leftrightarrow (\Psi\wedge\Phi) \vee (\Psi\wedge\Theta)$
- **De Morgan for $\wedge$:** $\neg(\Psi\wedge\Phi) \Leftrightarrow \neg\Psi\vee\neg \Phi$
- **Idempotence of $\vee$:** $\Psi\vee\Psi \Leftrightarrow \Psi$
- **Commutativity of $\vee$:** $\Psi\vee\Phi \Leftrightarrow \Phi\vee\Psi$
- **Associativity of $\vee$:** $\Psi \vee (\Phi\vee\Theta) \Leftrightarrow (\Psi\vee\Phi) \vee \Theta$
- **Distributivity of $\vee$ over $\wedge$:** $\Psi \vee (\Phi\wedge\Theta) \Leftrightarrow (\Psi\vee\Phi)\wedge (\Psi\vee\Theta)$
- **De Morgan for $\vee$:** $\neg(\Psi\vee\Phi) \Leftrightarrow \neg\Psi\wedge\neg\Phi$
- **Contrapositive for $\Rightarrow$:** $(\Psi\Rightarrow\Phi) \Leftrightarrow \neg\Psi\Rightarrow\neg\Phi$

## Index

**End Notes**

---

[1] Church's system was quickly shown to be inconsistent because it allowed Gödelian "self-referential" propositions, which lead to inconsistency in Mathematics [Hewitt 2011].

[2] [Kowalski1988]

[3] According to [Kowalski 1979]:

"*an inconsistent system can ... organize useful information... Thus* [finding a ] *contradiction, far from harming an information system, helps to indicate areas in which it can be improved. It facilitates the development of systems by successive approximation – daring conjectures followed by refutation and reconciliation. It favours bold, easily falsified beliefs, which can be weakened if need should arrive, over save, timid beliefs, which are difficult to strengthen later on. Better to make mistakes and to correct them than to make no progress at al.*"

[4] Raising issues of Inconsistency Robustness, The Obama administration deleted the following statement from its 2008 campaign website:

> "**Protect Whistleblowers**: Often the best source of information about waste, fraud, and abuse in government is an existing government employee committed to public integrity and willing to speak out. Such acts of courage and patriotism, which can sometimes save lives and often save taxpayer dollars, should be encouraged rather than stifled. We need to empower federal employees as watchdogs of wrongdoing and partners in performance. Barack Obama will strengthen whistleblower laws to protect federal workers who expose waste, fraud, and abuse of authority in government. Obama will ensure that federal agencies expedite the process for reviewing whistleblower claims and whistleblowers have full access to courts and due process."

It may be that Obama's Administration's statement on the importance of protecting whistleblowers went from being a promise for his administration to a political liability. There are manifest contradictions in what Obama said then and what he is doing now.

[5] In 1666, England's House of Commons introduced a bill against atheism and blasphemy, singling out Hobbes' Leviathan. Oxford university condemned and burnt Leviathan four years after the death of Hobbes in 1679.

6 In Latin, the principle is called ex falso quodlibet which means that from falsity anything follows.

7 [Nekham 1200, pp. 288-289]; later rediscovered and published in [Lewis and Langford 1932]

[8] [Pospesel 2000] has discussed extraneous ∨ introduction on in terms of the following principle: $\Psi$, $(\Psi\vee\Phi \vdash \Theta) \vdash \Theta$ However, the above principle immediately derives extraneous ∨ introduction when $\Theta$ is $\Psi\vee\Phi$. In Direct Logic, argumentation of the above form would often be reformulated as follows to eliminate the spurious $\Phi$ middle proposition: $\Psi$, $(\Psi \vdash \Theta) \vdash \Theta$

[9] [Green 1969].

[10] [Kowalski 1988]

[11] Resolution Theorem Proving is not Inconsistency Robust because it can be used to prove that there are no contradictions as follows:

Using Resolution Theorem Proving, $\neg(\Psi\wedge\neg\Psi)$ can be proved because

$\Psi,\neg\Psi \vdash_{ClassicalResolution}$ False

It is possible to use *Inconsistency-Robust Resolution* as follows:

$\Psi\vee\neg\Psi, \neg\Psi\vee\Phi, \Psi\vee\Omega \vdash_{T} \Phi\vee\Omega$ that requires the additional assumption $\Psi\vee\neg\Psi$ in order to make the inference.

Of course, it is possible to add the classical resolution rule to a theory $T$ by adding the following: $\Psi\vee\Phi, \neg\Phi\vee\Theta \vdash_{T} \Psi\vee\Theta$

[12] The argument below originated in [Nekham 1200, pp. 288-289] (later rediscovered and published in [Lewis and Langford 1932]) is an argument that an inconsistency $\vartheta\wedge\neg\vartheta$can be used to infer every proposition X. The Nekham argument is not valid in Inconsistency Robust Direct Logic because it make use of the rule of Extraneous ∨–Introduction, *i.e.*, $\Psi\vdash(\Psi\vee\Phi)$. The Nekham argument can be formalized is as follows: $\vartheta\vdash(\vartheta\vee X)$ and therefore $\vartheta,\neg\vartheta\vdash X$ because $\neg\vartheta,(\vartheta\vee X)\vdash X$

[13] [Wos et. al., 1965]

[14] I am grateful to Kowalski for clarifying his position. [Personal communication March 8, 2014].

[15] such as the one immediately above

[16] (later generalized, e.g., ActorScript [Hewitt 2013])

[17] A grounded-complete predicate is one for which all instances in which the predicate holds are explicitly manifest, i.e. instances can be generated using patterns. See [Ross and Sagiv 1992, Eisner and Filardo 2011].

[18] Execution can proceed differently depending on whether this set fits in memory.

[19] Execution can proceed differently depending on whether this set fits in memory.

[20] Sussman and Steele [1975] mistakenly concluded

> "*we discovered that the 'Actors' and the lambda expressions were identical in implementation*."

The actual situation is that the lambda calculus is capable of expressing some kinds of sequential and parallel control structures but, in general, not the concurrency expressed in the Actor model. On the other hand, the Actor model is capable of expressing everything in the lambda calculus and more.

Sussman and Steele noticed some similarities between Actor customers and continuations introduced by [Reynolds 1972] using a primitive called escape that was a further development of hairy control structure. In their program language Scheme, they called their variant of escape by the name "*call with current continuation*." Unfortunately, general use of escape is not compatible with usual hardware stack disciple introducing considerable operational inefficiency. Also, using escape can leave customers stranded. Consequently, use of escape is generally avoided these days and exceptions are used instead so that clean up can be performed. [Hewitt 2009]

[21] and unlike Prolog (see below)

[22] There was somewhat similar work that Hayes had discussed with the researchers at Aberdeen on ABSYS/ABSET [Foster and Elcock 1969].

[23] According to [Colmerauer and Roussel 1996]:

> *While attending an IJCAI convention in September '71 with Jean Trudel, we met Robert Kowalski again and heard a lecture by Terry Winograd on natural language processing. The fact that he did not use a unified formalism left us puzzled. It was at this time that we learned of the existence of Carl Hewitt's programming language, Planner [Hewitt, 1969]. The lack of formalization of this language, our ignorance of Lisp and, above all, the fact that we were absolutely devoted to logic meant that this work had little influence on our later research.*

However, according to [Kowalski 2008]:

> *During the next couple of years, I tried to reimplement Winograd's system in resolution logic and collaborated on this with Alain Colmerauer in Marseille. This led to the procedural interpretation of Horn clauses (Kowalski 1973/1974) and to Colmerauer's development of the programming language Prolog.*

[24] [Horn 1951]

[25] In practice, Prolog implemented a number of non-logical computational primitives for input-output, *etc.* Like Planner, for the sake of efficiency, it used backtracking. Prolog also had a non-logical computational primitive like the one of Planner to control backtracking by conditionally testing for the exhaustive failure to achieve a goal by backward chaining. However, Prolog was incapable of expressing strong "Negation as Failure" because it lacked both the assertions and true negation of Planner and thus it was impossible in Prolog to say "if attempting to achieve the goal G exhaustively

fails then assert (*not* G).” Prolog extended Planner by using unification (but not necessarily soundly because for efficiency reasons it can omit use of the “occurs” check).

[26] Prolog required a top-level goal $\Phi_1 \wedge \ldots \wedge \Phi_n$ to be stated as follows: **False**$\Leftarrow(\Phi_1 \wedge \ldots \wedge \Phi_n)$ [logically equivalent to the disjunctive clause $\neg\Phi_1 \vee \ldots \vee \neg\Phi_n$], which requires that in order to find solutions to the goal, the disjunctive clause must be refuted by deriving an contradiction, which is not Inconsistency Robust.

[27] some of material below was contributed by the author for publication in Wikipedia on “The Actor Model.”

[28] [Kowalski 1988a] On the other hand since the fall of 1972 with the invention of the Actor Model, Logic Programs can be rigorously defined very general terms (starting with the McCarthy’s Advice Taker proposal [McCarthy 1958]) as “*what can be programmed in mathematical logic*.” Of course, what can be programmed in mathematical logic is exactly “*each computational step* (*e.g.* as defined in the Actor model) *can be logically deduced*.” Even allowing the full power of Direct Logic, computation is not reducible to Logic Programs [Hewitt 2011].

[29] *cf.* Plotkin [1976]

[30] When asserted that aResult is sent in *aResponse*

[31] Assert in Unbounded1[*aRequest*] that **Counter** is sent [ ]

[32] e.g. [Kowalski and Sadri 2015]

[33] ICOT used monotonic *mutable* lists instead of events in its Prolog-style clause programs.

[34] later adapted for concurrency, e.g., Java, etc.

[35] Use of multiple ports is an awkward programming idiom that introduces many difficulties, *e.g.*, starvation due not properly servicing a port. For a contrary view, see [Kahn and Saraswat 1990].

[36] **Postpone** *expression* delays execution of *expression* until the result is needed.

[37] Execution can proceed differently depending on whether this set fits in memory.

[38] [Kowalski 2014]

[39] By the *Computational Representation Theorem* [Clinger 1981; Hewitt 2006], which can define all the possible executions of a procedure.

[40] e.g. [Shulman 2012, nLab 2014]

[41] ℕ is the type of Natural Numbers.

[42] type of 2-element list with first element of type $\sigma_1$ and with second element of type $\sigma_2$

[43] type of computable procedures from type $\sigma_1$ into $\sigma_2$.

[44] type of functions from $\sigma 1$ into $\sigma 2$

[45] type of term of type $\boldsymbol{\sigma}$

[46] *if* **t** *then* $\Phi_1$ *else* $\Phi_2$

[47] $\Phi_1$, … and $\Phi_k$ infer $\Psi_1$, …, and $\Psi_n$

[48] **p** is a proof of $\Phi$

[49] *if* $\mathbf{t}_1$ *then* $\mathbf{t}_2$ *else* $\mathbf{t}_3$

[50] Because there is no type restriction, fixed points may be freely used to define recursive procedures on expressions.

[51] *if* **t** *then* $\mathbf{s}_1$ *else* $\mathbf{s}_1$